\documentclass[12pt]{article}
\newlength{\vshift}
\input epsf.tex
\newlength{\hshift}

\usepackage{amssymb}
\usepackage{amsmath}
\usepackage{amsthm}
\usepackage{amsopn}
\def\beq{\begin{equation}}
\def\eeq{\end{equation}}
\def\bea{\begin{eqnarray}}
\def\eea{\end{eqnarray}}

\def\d{\delta}
\def\ep{\epsilon}
\def\no{\nonumber}
\def\uno{\mbox{1 \kern-.59em {\rm l}}}
 
\begin{document}

 \vspace*{3cm}

\begin{center}

{ \Large \bf{Generation of circular polarization of the CMB}}

\vskip 4em

 {\bf E. Bavarsad},
{\bf M. Haghighat}, {\bf R. Mohammadi}, {\bf I. Motie}, {\bf Z. Rezaei}  \\ and {{\bf M. Zarei}} 

\vskip 1em

Department of Physics, Isfahan University of Technology, Isfahan
84156-83111, Iran
 \end{center}

 \vspace*{1.5cm}

\begin{abstract}
According to the standard cosmology, near the last scattering surface, the photons scattered via Compton scattering are just linearly polarized and then the primordial circular polarization of the CMB photons is zero. In this work
we show that CMB polarization acquires a small degree of circular polarization when a background magnetic field is considered or the quantum electrodynamic sector of standard model is extended by Lorentz-noninvariant operators as well as noncommutativity.  The existence of circular polarization for the CMB radiation may be verified during
future observation programs and it represents a possible new channel for investigating new physics effects.

\end{abstract}
%%%%%%%%%%%%%%%%%%%%%%%%%%%%%%%%%%%%%%%%%%%%%%%%%%%%%%%%%%%%%%%%%%%%%%%%%%%%%%%%%%%%%%%%%%%%%%%%%%%%%%%%%%%%%%%%%%%%%%%%%%%
\section{ Introduction}

The observation of polarization in the cosmic microwave background
(CMB) and its correlation to the temperature anisotropies is a
very promising tool to test the physics of early universe. The CMB
radiation is expected to be linearly  polarized of the order of  10 \%. This
Linearity is a result of the anisotropic Compton scattering around
the epoch of recombination which has been widely discussed in the
literature \cite{Zaldarriaga:1996xe,Hu:1997hv,Kosowsky:1994cy}. A linearly polarized radiation is described
by non-zero values for the Stokes parameters
$Q$ and/or $U$. Meanwhile the possibility of the generation of circular
polarization can be determined by the Stokes
parameter $V$. This parameter is usually given to be zero in the
literature because there is no physical mechanism to
generate a Stokes-V contribution at the last scattering surface. However the circular polarization can be generated naturally if a background magnetic field be present. Recently, Giovannini has shown that if the CMB photons are scattered via electrons in the
presence of a magnetic field, a non-vanishing $V$ mode can be produced \cite{Giovannini1,Giovannini2}.
Furthermore, Cooray,
Melchiorri and Silk have discussed that the CMB radiation observed
today is not exactly the same as the field last scattered
\cite{Cooray:2002nm}. The CMB polarization may change during the
propagation from the last scattering surface due to the presence of
relativistic magnetized plasma in galaxy clusters. On the
basis of the mechanism described in \cite{Cooray:2002nm}, the linear polarization of
the CMB can be converted to the circular polarization under the
formalism of the generalized Faraday rotation \cite{faraday1}
known as the Faraday conversion \cite{faraday2}. The evolution of the
Stokes parameter $V$ given by this mechanism is obtained as \beq
\dot{V}=2\:U\frac{d\Delta\phi_{FC}}{dt}, \eeq
where
$\Delta\phi_{FC}$ is the Faraday conversion phase shift \cite{Cooray:2002nm}. Assuming reasonable parameters for the relativistic
magnetized plasma in the galaxy clusters and for $B=10\:\mu G$, one can estimate the Faraday conversion phase shift as $\Delta\phi_{FC}\sim \texttt{few} \times 10^{-3}$. Then with a CMB polarization of the order of $10^{-6}$ which propagates through the galaxy clusters, the outgoing beam should contain a circular polarization of order $10^{-9}$ at scales corresponding to the galaxy clusters \cite{Cooray:2002nm}.

It is also possible that such an effect arise due to new physics effects. For instance
another mechanism for generating the circular polarization
has been reported by Alexander, Ochoa and Kosowsky in
\cite{kosowsky}. In this method the photon sector of quantum
electrodynamics is extended so that photon is coupled to an
external constant 4-vector $T_{\nu}$ via a Chern-Simons like term as
\bea
\mathcal{L}'\!\!\!\!\!\!\!\!\!&&=\mathcal{L}_{Maxwell}+\mathcal{L}_{T}
\nonumber \\&&
=-\frac{1}{4}F_{\mu\nu}F^{\mu\nu}+g\epsilon^{\mu\nu\alpha\beta}A_{\mu}T_{\nu}
F_{\alpha\beta},\label{lo} \eea
where $\mathcal{L}_{T}$ is a Lorentz symmetry violation CPT odd
term and $g$ is the coupling constant of interaction. Then the collision term on the right hand side of
the Boltzmann equation is corrected
by the interaction (\ref{lo}). The generalized Boltzmann
equation considered in this reference is given by the
quantum-mechanical evolution of the photon density matrix. In fact it is shown that
the time derivative of the polarization brightness associated with the Stokes parameter $V$ receives a source term such that the $V$ mode becomes nonzero.

Also in reference \cite{axion},
 Finellia and Galavernid have
considered an axion-like cosmological pseudoscalar field acting as dark matter coupled to
the photons as
  \beq \mathcal{L}=-\frac{g_{\phi}}{4}\phi
F_{\mu\nu}\tilde{F}^{\mu\nu}, \eeq

where $g_{\phi}$ is the coupling
constant between axion field $\phi$ and the electromagnetic field
strength $F_{\mu\nu}$ and
$\tilde{F}^{\rho\sigma}=\epsilon^{\mu\nu\rho\sigma}F_{\rho\sigma}$.
Then it has been shown that such an interaction between
the pseudoscalar field and photons rotates the plane
of linear polarization and generates the circular polarization for the cosmic microwave background.
A similar approach with the pseudoscalar photon mixing shows a CMB circular polarization up to the order of $10^{-7}$ \cite{Agarwal:2008ac}.

In this work we address new possibilities and propose new channels to produce the circular polarization for the CMB in three different contexts.  First we consider the QED part of an effective field theory for Lorentz violation (LV) symmetry \cite{l,l0,l1,l2,l3,l4,l5} that is called the Standard-Model Extension (SME) \cite{l0}. In this case we only investigate that Lorentz violation terms of SME that have not been considered in \cite{kosowsky}. Then we examine the noncommutative QED (NCQED) with Seiberg-Witten expansion of fields \cite{nc} as well as the presence of the primordial magnetic field background in the last scattering surface \cite{magnetic}.
 For this purpose we use the generalized Boltzmann equation formulated in \cite{kosowsky}, to study the evolution of Stokes parameter $V$ under the influence of these new phenomena.  This equation shows that a non-zero background field can generate the circular polarization through either the interaction of photon with the background itself or the corrections on the Compton scattering of photon on the electron in the presence of the background field.  One should note that the quantum corrections on the scattering cross section, for the low energy particles, are usually negligible. Nevertheless, the circular polarization of CMB itself as a small quantity can be produced potentially via accumulation of these small effects.

On the dimensional ground one can expect $\Delta\phi_{FC}$ at the lowest order depends on the appropriate combination of the parameters of the model which is dimensionless.  Therefore, for the magnetic field $\Delta\phi_{FC}$ should be proportional to $eB/m^2$ that is about $10^{-19}$ for the micro gauss magnetic field.  Meanwhile in the NC space it should be depended on  $\alpha m^2\theta$ which is of the order $10^{-17}$ for $1/\sqrt{\theta}$ about 10 TeV.  For the Lorentz violation case the Faraday conversion phase should be linearly proportional to the dimensionless LV-parameters which are less than the present existing bound about $10^{-15}$.
 Therefore at first sight it seems that the effect of the magnetic field is at least two order of magnitudes less than the other new interactions.  But as we will show soon this is not the case and due to the wavelength dependence of the solutions the background magnetic field has main contribution on the polarization of the CMB.  However, it should  be also noted that the resulting circular polarization modes are certainly small due to the smallness of quantum effects. Meanwhile the probe of circular polarization of the CMB radiation through the future subtle experiments, certainly give more information about the new physics or the primordial magnetic fields in the early universe.
 The main condition which is assumed everywhere during the paper is a cold plasma at the last scattering surface at a red shift of about $1100$ or an age of about $180, 000 \:(\Omega_{0}h^{2})^{-1/2} \:\texttt{yrs}$ and a temperature of the order of $ \mathcal{O}(\texttt{eV})$. The low temperature is always applicable for the calculation of the Compton scattering amplitude.

The paper is organized as follows: In section 2 we review the Stokes parameters formalism and the generalized Boltzmann equation. In section 3 the generation of circular polarization due to the Lorentz violation is discussed. In section 4 the same calculation is done for the noncommutativity.  In  section 5 we examine  the generation of circular polarization in the presence of a background magnetic field. Finally in the last section we discuss about the result.

 %%%%%%%%%%%%%%%%%%%%%%%%

 \section{Stokes parameters and Boltzmann equation}

The polarization of CMB is usually characterized by means of the
Stokes parameters of radiation: I, Q, U and V \cite{Kosowsky:1994cy}. For a
quasi-monochromatic wave propagating in the z-direction, in which
the electric and magnetic fields vibrate on the x-y plane, the
electric field $E$ in a given point can be written as
\beq E_x = a_x(t)\cos(\omega_0t-\delta_x)
,\:\:\:\:\:\:\:\:\:\:\:\: E_y = a_y(t) \cos (\omega_0t-
\delta_y),\label{ee} \eeq
where the wave is nearly monochromatic with frequency $\omega_{0}$.
Then the Stokes parameters are defined by time averaging of the
parameters of electric fields (\ref{ee}) as follows
\bea &&I\equiv\left<a_x^2\right>+\left<a_y^2\right>,\\
&&Q\equiv\left<a_x^2\right>-\left<a_y^2\right>,\\
&&U\equiv\left<a_xa_y\cos(\delta_x-\delta_y) \right>,\\
&&V\equiv\left<a_xa_y\sin(\delta_x-\delta_y) \right>.
 \eea
 These parameters are physically interpreted as follows: I is the total intensity of wave, Q measures the difference between $x$ and $y$ polarization, the parameter U gives the phase information for the two linear polarization and V determines the difference between positive and negative circular polarization.
 The time evolution of these Stokes parameters is given through Boltzmann equation.
 We now turn to the Boltzmann equation which is vital in our study. The CMB radiation is
generally described by the phase space distribution function $f$
for each polarization state. Boltzmann equation is a systematic
mechanism in order to describe the evolution of the distribution
function under gravity and collisions. The classical Boltzmann
equation generally is written as \cite{book0}
 \beq \frac{df}{dt}=C[f], \eeq
where the left hand side is known as the Liouville term deals with the effects of gravitational perturbations about the homogeneous cosmology. An important point that must be mentioned here is that we have disregarded the scalar fluctuation of metric on the left hand side of the  Boltzmann equation.
 The right hand side of the
Boltzmann equation contains all possible collision terms. For
photons, the Compton scattering with free electrons, $
\gamma(p)+e(q)\leftrightarrow \gamma(p')+e(q')$,
 is more important and must be included on the right hand side. To compute the evolution of polarization we follow
 the approach of references \cite{Kosowsky:1994cy,kosowsky}. At first, the distribution function $f$ is generalized to the density matrix
 $\rho_{ij}$
  \bea \rho\!\!\!\!\!\!&&=\frac{1}{2}\left(
                       \begin{array}{cccc}
                         I+Q & U-i V \\
                          U+iV & I-Q  \\
                        \end{array}
                     \right)\no \\&&
                     =\frac{1}{2}\:(I\uno+Q\sigma_3+U\sigma_1+V\sigma_2),\label{matrix} \eea
 where $\uno$ is the identity matrix and $\sigma_i$ are the Pauli spin matrices. The density matrix $\rho_{ij}$ is related to the number operator $\mathcal{\hat D}_{ij}({\bf k}) = \hat a_i^\dagger({\bf k})\hat a_j ({\bf k})$ as
 \beq \label{eq:3.17}
        \langle \mathcal{\hat D}_{ij}({\bf k}) \rangle = (2\pi )^3 2k^0\delta^{(3)}(0)\rho_{ij}({\bf k}).\eeq
and the time evolution of the number operator is given by

 \begin{equation}
\label{eq:3.26}
        \left\langle \frac{d}{dt}\mathcal{\hat D}_{ij} \right\rangle(t) \simeq i\langle [ \mathcal{\hat H}_{\text{int}}(t) , \mathcal{\hat D}_{ij} ] \rangle -\int_0^t dt' \langle [\mathcal{\hat H}_{\text{int}}(t-t'),[ \mathcal{\hat H}_{\text{int}}(t') , \mathcal{\hat D}_{ij} ] ]\rangle .
\end{equation}
Hence in terms of $\rho$, the evolution equation leads to
\begin{equation}
\label{gb0}
        (2\pi )^3 \delta^3 (0)2k^0 \frac{d} {dt} \rho_{ij}(0,{\bf k})=i\langle[\mathcal{\hat H}_{\text{int}}(0),
        \mathcal{\hat D}_{ij}({\bf k})]\rangle -  \frac{1}{2}\int_{-\infty}^\infty dt'\phantom{1}
        \langle[  \mathcal{\hat H}_{\text{int}}(t'),[\mathcal{\hat H}_{\text{int}}(0),\mathcal{\hat D}_{ij}({\bf k})]]\rangle,
\end{equation}
where the first term on the right hand side  is referred as the refractive term and the second one as the damping term.
Note that in (\ref{gb0}), all factors of $\delta^3(0)$ will be
canceled from the final expressions. Equation (\ref{gb0}) can be
viewed as a quantum mechanical Boltzmann equation for the phase
space function $\rho$. In the case where the interaction term is
the convenient QED interaction, the time evolution of the Stokes
parameter $V$ is always equal to zero which leads to the absence
of circular degrees of freedom for the CMB photons. In this
paper during the next sections, we will show that if one
consider beyond standard model interaction terms, the time
evolution of the $V$ receives a source term which is interpreted as a
non-zero circular polarization. If the Stokes parameter $V$ be
non-zero, then it contributes on  the new angular power
spectrum elements such as $C_{l}^{VT}$, $C_{l}^{VE}$ and $C_{l}^{VV}$
 where may be detectable in future CMB experiments.

%%%%%%%%%%%%%%%%%%%%%%%%%%%%%%%%%%%%%%%%%%%%%%%%%%%%%%%%%%%%%%%%%%%%%%%%%%%%%%%%%%%%%%%%%%%
\section{Generation of circular polarization in the presence of Lorentz violation terms}

In the present section we consider a new class of Lorentz invariance violation (LIV) terms as a generic class of
interaction between photons and an external field and explore whether they can
produce circular polarization for the CMB photons or not. Although the source of these asymmetry terms is unknown,
physicists usually suppose that they come from presumably short-distance
physics which manifest themselves in the interactions of standard
model (SM) fields \cite{l,l0,l1,l2,l3,l4,l5}. The renormalizable sector of the general QED extension for a single Dirac field $\psi$ of mass $m$ and a photon also has been studied by some authors for which the Lagrangian $\mathcal{L}$ is written as
\beq \mathcal{L} =\mathcal{L}_{\rm QED}+
\mathcal{L}_{LIV}^{electron}+\mathcal{L}_{LIV}^{photon},\eeq
where
 \bea
\mathcal{L}_{\rm QED}=i\: \bar{\psi} \gamma^\mu \mathcal{D}_{\mu}
\psi - m
\bar{\psi}\psi -\frac{1}{4}F_{\mu\nu}F^{\mu\nu},\\
\mathcal{L}_{LIV}^{electron}=
\frac{i}{2}\:c_{\mu\nu}\bar{\psi}\gamma^{\mu}\mathcal{D}^{\nu}\psi
+\frac{i}{2}\:d_{\mu\nu}\bar{\psi}\gamma_{5}\gamma^{\mu}\mathcal{D}^{\nu}\psi\label{cd},\\
\mathcal{L}_{LIV}^{photon}=
-\frac{1}{4}\:(k_{F})_{\mu\nu\alpha\beta}F^{\alpha\beta}F^{\mu\nu}+
\frac{1}{2}(k_{AF})^{\alpha}\epsilon_{\alpha\beta\mu\nu}A^{\beta}F^{\mu\nu},\label{ext}
\eea
where the $\mathcal{D}_{\mu}$ is the covariant derivative.
 In the electron sector,
$c_{\mu\nu}$ and $d_{\mu\nu}$ are the dimensionless Hermitian coefficients with
symmetric and antisymmetric space-time components. For the photon sector, it will be useful to decompose the coefficient $k_F$
into a tensor with 10 independent
components analogous to the Weyl tensor in
general relativity and one with 9 components analogous
to the trace-free Ricci tensor with the following symmetries of the Riemann tensor
 \beq \label{symm}
 \:(k_{F})_{\mu\nu\alpha\beta}=
 -\:(k_{F})_{\nu\mu\alpha\beta}=-\:(k_{F})_{\mu\nu\beta\alpha}=\:(k_{F})_{\nu\mu\beta\alpha}=\:(k_{F})_{\alpha\beta\mu\nu}.\eeq
So it
contains 19 independent real components.
 Various
experiments have bounded these coefficients both in fermion and photon sectors. For instance, the only rotation-
invariant component of $(k_{F})_{\mu\nu\beta\alpha}$ is constrained to $\leq 10^{-23}$ \cite{l} and
all other components of
$(k_{F})_{\mu\nu\beta\alpha}$ which are associated with violations of rotation invariance can be bounded of about
$\leq 10^{-27}$ \cite{l0}.
 The high-quality spectropolarimetery of distant
galaxies at infrared, optical, and ultraviolet frequencies, also constrains   the coefficients $k_F$ to less than $3\times 10^{-32}$ \cite{l2}. In references \cite{l3,l4}, another bound has been estimated. For the couplings $c_{\mu\nu}$ and $d_{\mu\nu}$ in the electron sector, one can find a variety of bounds such as $c_{\mu\nu}\leq 10^{-15}$ and $d_{\mu\nu}\leq 10^{-14}$ in \cite{l5} and references there in.

 The $\frac{1}{2}(k_{AF})^{\alpha}\epsilon_{\alpha\beta\mu\nu}A^{\beta}F^{\mu\nu}$ term in the photon sector, has been studied in \cite{kosowsky} and it was shown that the CMB photons acquire  small degree of circular polarization. In this section we consider the three remaining terms in the LIV Lagrangian.
 We study the contributions of each terms in the Boltzmann equation and calculate
the evolution of the Stokes parameters. Firstly, the photon sector term with $k_F$ coupling and then the electron sector with $c_{\mu\nu}$ and $d_{\mu\nu}$ couplings are investigated.

\subsection{ The pure photon sector }

In this part the Lorentz violation terms (\ref{ext}) are investigated. These terms do not modify the Compton scattering but the dynamics of the CMB photons can be influenced. Here we consider $\mathcal{\hat{H}}_{\text{int}}=-\frac{1}{4}\:(k_{F})_{\mu\nu\alpha\beta}F^{\alpha\beta}F^{\mu\nu}$  on the right hand side of the generalized Boltzmann equation (\ref{gb0}).
Using the canonical commutation relations of the
creation and annihilation operators and their expectation values given in \cite{Kosowsky:1994cy}, the equations (4.11a)-(4.11e), one can calculate the refractive term as
\begin{eqnarray}
    \label{eq:4.44}
        \left\langle\left[\mathcal{ \hat H}_{\text{int}}(t),\mathcal{\hat D}_{ij}({\bf k})\right]
        \right\rangle = -\frac{1}{2}\:(k_{F})^{\mu\nu\alpha\beta}(2\pi )^3 \delta
        ^{(3)}(0)(k_{\alpha}k_{\mu} \epsilon^* _{s'\beta} \epsilon
        _{s\nu})\left[ \delta_{si}\rho_{s'j}( {\bf k})-\delta_{js'}\rho_{is}({\bf k}) \right
        ],
\end{eqnarray}
where $\epsilon _{s\mu}(k)$ are the photon polarization 4-vectors
with $s=1,2$. Now using
the symmetry properties of $\:k_{F}^{\mu\nu\alpha\beta}$ summarized in Eq. (\ref{symm}), one can write $\dot{\rho}_{ij}$ as
\begin{eqnarray}
    \label{eq:12}
        \frac{d} {dt} \rho_{ij}({\bf k})=-
        \frac{4}{k^{0}}\:(k_{F})^{\mu\nu\alpha\beta}(k_{\alpha}k_{\mu} \epsilon _{s'\beta} \epsilon^*
        _{s\nu})\left[ \delta_{s'i}\rho_{sj}( {\bf k})-\delta_{js'}\rho_{is}({\bf k}) \right
        ].
\end{eqnarray}
The photon
density matrix can be expanded about a uniform unpolarized distribution $\rho^{(0)}$ as follows
\beq \rho_{ij}=\rho_{ij}^{(0)} +\rho_{ij}^{(1)},\eeq
where $\rho_{11}^{(0)}=\rho_{22}^{(0)}$ and $\rho_{12}^{(0)}=\rho_{21}^{(0)}=0$.
Therefore the components of the time derivative of the density matrix are given by
\begin{eqnarray}
    \label{eq:}
        \dot{\rho}_{11}^{(1)}({\bf k})=
        -\frac{4}{\:k^{0}}\:(k_{F})^{\mu\nu\alpha\beta}k_{\alpha}k_{\mu}[\epsilon _{1\beta} \epsilon^*
        _{2\nu}\rho_{21}^{(1)}-\epsilon _{2\beta} \epsilon^*
        _{1\nu}\rho_{12}^{(1)}],
\end{eqnarray}
\begin{eqnarray}
    \label{eq:}
       \dot{\rho}_{22}^{(1)}({\bf k})=
         -\frac{4}{\:k^{0}}\:(k_{F})^{\mu\nu\alpha\beta}k_{\alpha}k_{\mu}[\epsilon _{2\beta} \epsilon^*
        _{1\nu}\rho_{12}^{(1)}-\epsilon _{1\beta} \epsilon^*
        _{2\nu}\rho_{21}^{(1)}],
\end{eqnarray}
\begin{eqnarray}
    \label{eq:}
        \dot{\rho}_{12}^{(1)}({\bf k})=
        -\frac{4}{\:k^{0}}\:(k_{F})^{\mu\nu\alpha\beta}k_{\alpha}k_{\mu}[\epsilon _{1\beta} \epsilon^*
        _{1\nu}\rho_{12}^{(1)}+\epsilon _{1\beta} \epsilon^*
        _{2\nu}\rho_{22}^{(1)}-\epsilon _{1\beta} \epsilon^*
        _{2\nu}\rho_{11}^{(1)}-\epsilon _{2\beta} \epsilon^*
        _{2\nu}\rho_{12}^{(1)}],
\end{eqnarray}
\begin{eqnarray}
    \label{eq:}
        \dot{\rho}_{21}^{(1)}({\bf k})= -\frac{4}{\:k^{0}}\:
        (k_{F})^{\mu\nu\alpha\beta}k_{\alpha}k_{\mu}[\epsilon _{2\beta} \epsilon^*
        _{1\nu}\rho_{11}^{(1)}+\epsilon _{2\beta} \epsilon^*
        _{2\nu}\rho_{21}^{(1)}-\epsilon _{1\beta} \epsilon^*
        _{1\nu}\rho_{21}^{(1)}-\epsilon _{2\beta} \epsilon^*
        _{1\nu}\rho_{22}^{(1)}].
\end{eqnarray}

Note that the polarization vectors $\epsilon _{s\mu}(q)$ are
real. Hence the evaluation of the Stokes parameters are given as
\beq
    \label{eq:}
        \dot{I}^{(1)}=0,
      \eeq

          \beq \dot{Q}^{(1)}=-\frac{16}{k^{0}}\:(k_{F})^{\mu\nu\alpha\beta}k_{\alpha}k_{\mu}(\epsilon _{1\beta} \epsilon^*
        _{2\nu}-\epsilon _{2\beta} \epsilon^*
        _{1\nu})\:V^{(1)},
    \label{eq:}
    \eeq

    \beq
        \dot{U}^{(1)}= -\frac{4}{k^{0}}\:(k_{F})^{\mu\nu\alpha\beta}k_{\alpha}k_{\mu}(\epsilon _{2\beta} \epsilon^*
        _{1\nu}-\epsilon _{1\beta} \epsilon^*
        _{2\nu})\:Q^{(1)},
    \label{eq:}
    \eeq

     \beq
        \dot{V}^{(1)}=-\frac{4}{k^{0}}\:(k_{F})^{\mu\nu\alpha\beta}k_{\alpha}k_{\mu}\Big\{(\epsilon _{1\beta} \epsilon^*
        _{1\nu}-\epsilon _{2\beta} \epsilon^*
        _{2\nu})\:U^{(1)}-(\epsilon _{1\beta} \epsilon^*
        _{2\nu}+\epsilon _{2\beta} \epsilon^*
        _{1\nu})\:Q^{(1)}\Big\}.\label{eq:29}
\eeq

Therefore in the first order of Lorentz violation parameter $k_{F}$, $\dot{V} $ is nonzero. This shows that although the usual Compton scattering of CMB photons can't generate circular polarization, the interaction of photon with a nontrivial LIV background clearly produces a nonzero circular polarization.  To have an estimate of the Faraday phase conversion we rewrite (\ref{eq:29}) as follows
\beq
        \dot{V}^{(1)}=-4k^{0}\:(k_{F})^{\mu\nu\alpha\beta}\Big\{L_{\alpha\mu\beta\nu}^U\:U^{(1)}-L_{\alpha\mu\beta\nu}^Q\:Q^{(1)}\Big\},\label{LVP}
\eeq
where $L_{\alpha\mu\beta\nu}^U=\frac{k_{\alpha}}{k^{0}}\frac{k_{\mu}}{k^{0}}(\epsilon _{1\beta} \epsilon^*_{1\nu}-\epsilon _{2\beta} \epsilon^*_{2\nu})$ and a corresponding relation for $L_{\alpha\mu\beta\nu}^Q$.  Now for $\lambda_0=1cm$, z=1000 and for 1 kpc one has

\beq
\Delta\phi_{FC}\sim 4\times10^{20}k_F,
\eeq
or to have a phase of order one we should have $k_F\sim 10^{-20}$.

%%%%%%%%%%%%%%%%%%%%%%%%%%%%%%%%%%%%%%%%%%%%%%%%%%%%%%%%%%%%%%%%%%%%%%%%%%%%%%%%%%%%%%%%%%%%%%%%%%%%%%%%%%

\subsection{The electron sector }
\subsubsection{$c^{\mu\nu}$ term }

We now turn to the LIV terms in the electron sector (\ref{cd}). Calculating the refractive term on the right hand side of the generalized Boltzmann equation (\ref{gb0}), leads to $\dot{V}=0$ for interaction (\ref{cd}).
Therefore, We move on to the damping term. This term gives corrections to the Compton scattering of the CMB photons and electrons.
 These corrections modify the right hand side of the equation (\ref{gb0}) and then may influence the evolution of the Stokes parameter $V$. In this part we investigate whether these corrections lead to a circular polarization mode or not. Firstly, the gauge field $A_{\mu}$ is expanded in terms of annihilation and creation operators $a$ and $a^{\dag}$ and the fermion fields $\psi$ and $\bar{\psi}$ are expended in terms of $b$ and $b^{\dag}$. Then
the interaction
Hamiltonian can be written in the momentum space as
\begin{eqnarray}
    \label{eq:Hint.}
   \mathcal{\hat H}_{int}(t)&&\!\!\!\!\!\!\!\!= \int d {\bf q} \,d {\bf q'}\,  d{\bf p} \,  d{\bf p'} \,
         \:(2\pi )^3 \delta^{(3)}({\bf q'}+{\bf p'}-{\bf q}-{\bf
         p})\exp(it(q'^0+p'^0-q^o-p^0))\no \\&&\:\:\:\:\:\:\:\times\: b^{\dag}_{r'}(q')a^{\dag}_{s'}(p')({\cal {M}}_1+{\cal {M}}_2)a_s(p)b_r(q),
\end{eqnarray}
where $q$ and $q'$ are the momenta of electrons and $p$ and $p'$ are the momenta of photons.
 Also we have used the following shorthand notations
\begin{eqnarray}
    \label{eq:abb.}
    d {\bf q}=\frac{d^3 {\bf q}}{(2\pi )^3}\frac{m}{q_0},\:\:\:\:\:\:\:\:\:\:\:\:d {\bf p}=\frac{d^3 {\bf p}}{(2\pi )^32p^0}.
\end{eqnarray}
The ${\cal {M}}_1$ and ${\cal {M}}_2$ are the amplitude for the Compton scattering process up to the first order of LIV parameters $c^{\mu\nu}$ and $d^{\mu\nu}$.
Consequently, the commutator in the damping term, finds the following form
\begin{eqnarray}
    \label{eqcomm}
    \left[\mathcal{\hat H}_{\text{int}}(t),\mathcal{\hat D}_{ij}({\bf
    k})\right]&&\!\!\!\!\!\!\!= \int d {\bf q} \,d {\bf q'}\,  d{\bf p} \,  d{\bf p'} \,
         \:(2\pi )^3 \delta^{(3)}({\bf q'}+{\bf p'}-{\bf q}-{\bf
         p})({\cal {M}}_1+{\cal {M}}_2)\no\\&&\!\!\!\! \times[b^{\dag}_{r'}(q')b_r(q)a^{\dag}_{s'}(p')a_s(p)2p^0(2\pi )^3
         \delta_{is}\delta^3({\bf p}-{\bf k})\no\\&&
        \:\:\:\:\:\:\:\:\:\:\:\:\:\:\:\:\:\:\: - b^{\dag}_{r'}(q')b_r(q)a^{\dag}_{s'}(p')a_s(p)2p'^0(2\pi )^3
         \delta_{js'}\delta^3({\bf p'}-{\bf k})].
\end{eqnarray}
In appendix A we detail the calculation of (\ref{eqcomm}) and it's expectation value. Quoting the results of Appendix A,
and expanding the density matrix as
$\rho_{ij}=\rho^{(0)}_{ij}+\rho^{(1)}_{ij}$,  the evolution of Stokes parameters is derived as
\begin{eqnarray}
    \label{eq:}
      \dot{I}^{(1)} = 0,
\end{eqnarray}

\bea \label{eq:}
        \dot{Q}^{(1)}&&\!\!\!\!\!\!=\frac{e^{2}}{2mk^{0}}\int {d{\bf
        q}}\:n_{e}({\bf q})\Big[\:
        \frac{2c^{\mu\nu}}{q.k}\{q.k\epsilon_{\mu}^{1}\epsilon_{\nu}^{2}-k_{\nu}q.\epsilon^{2}
        \epsilon_{\mu}^{1}-q.k\epsilon_{\nu}^{1}\epsilon_{\mu}^{2}+k_{\nu}q.\epsilon^{1}\epsilon_{\mu}^{2}\nonumber\\&&
       \:\:\:\:\:\:\:\:\:\:\:\:\:\: + \:q_{\nu}(q.\epsilon^{1}\epsilon_{\mu}^{2}+q.\epsilon^{2}\epsilon_{\mu}^{1})\}        +\frac{4c^{\mu\nu}}{(q.k)^{2}}(q_{\nu}q_{\mu}+k_{\nu}k_{\mu})(q.\epsilon^{1}q.\epsilon^{2})
        \:\Big]V^{(1)},
                    \eea

\begin{eqnarray}
    \label{eq:}
        \dot{U}^{(1)}&&\!\!\!\!\!\!=\frac{e^{2}}{2k^{0}m}\int {d{\bf
        q}}n_{e}({\bf
        q})\Big[\frac{2c^{\mu\nu}}{q.k}
        \{q.k(\epsilon_{\mu}^{1}\epsilon_{\nu}^{1}-\epsilon_{\mu}^{2}
        \epsilon_{\nu}^{2})-k_{\nu}q.\epsilon^{1}\epsilon_{\mu}^{1}+k_{\nu}q.\epsilon^{2}\epsilon_{\mu}^{2} \nonumber\\&&+
         2q_{\nu}(q.\epsilon^{1}
         \epsilon_{\mu}^{1}-q.\epsilon^{2}\epsilon_{\mu}^{2}) \}
                 +\frac{2c^{\mu\nu}}{(q.k)^{2}}(q_{\nu}q_{\mu}+k_{\nu}k_{\mu})(q.\epsilon^{1}
        q.\epsilon^{1}-q.\epsilon^{2}q.\epsilon^{2})\Big]V^{(1)},
 \end{eqnarray}
and
 \begin{eqnarray}
    \label{eq:}
        \dot{V}^{(1)} &&\!\!\!\!\!\!=\frac{ie^{2}}{2k^{0}m}\int {d{\bf
        q}}n_{e}({\bf
        q})\Big\{\Big(\frac{2c^{\mu\nu}}{q.k}\{(q.k(\epsilon_{\mu}^{1}
        \epsilon_{\nu}^{2}+\epsilon_{\mu}^{2}\epsilon_{\nu}^{1})-k_{\nu}q.
        \epsilon^{2}\epsilon_{\mu}^{1}+k_{\nu}q.\epsilon^{1}\epsilon_{\mu}^{2} \nonumber\\&&
        +q_{\nu}(q.\epsilon^{1}
        \epsilon_{\mu}^{2}+q.\epsilon^{2}\epsilon_{\mu}^{1})\} +\frac{c^{\mu\nu}}{(q.k)^{2}}(q_{\nu}q_{\mu}+k_{\nu}k_{\mu})(q.\epsilon^{1}q.\epsilon^{2})\Big)(-Q^{(1)})\nonumber\\&&+
       \Big(\frac{c^{\mu\nu}}{q.k}\{q.k(\epsilon_{\mu}^{1}\epsilon_{\nu}^{1}-\epsilon_{\mu}^{2}\epsilon_{\nu}^{2})
       -k_{\nu}q.\epsilon^{1}\epsilon_{\mu}^{1}+k_{\nu}
        q.\epsilon^{2}\epsilon_{\mu}^{2} \nonumber\\&& +q_{\nu}(q.\epsilon^{1}
        \epsilon_{\mu}^{1}-q.\epsilon^{2}\epsilon_{\mu}^{2})\}
     -        \frac{2c^{\mu\nu}}{(q.k)^{2}}(q_{\nu}q_{\mu}+k_{\nu}k_{\mu})
       (q.\epsilon^{1}q.\epsilon^{1}-q.\epsilon^{2}q.\epsilon^{2})\:\Big )U^{(1)}
        \Big\}.
\end{eqnarray}
An exact solution for the evolution of the Stokes parameters involving the coupled evolution equation for all of the species can only be accomplished through numerical integration.
The time derivative of the Stokes parameter $V$ has a source term which indicates the appearance of circular polarization for the CMB photons if they scatter from electrons through Compton process with an extra Lorentz violation $c_{\mu\nu}$ term.  Here again to have an estimate of the value of circular polarization we can find the order of magnitude of the Faraday phase conversion.  For this purpose we consider only the largest term and confine ourselves to $q_0\simeq m \gg |\bf q| , |\bf k|$.  Therefore one has
\begin{eqnarray}
    \label{eq:LVC}
        \dot{V}^{(1)} &&\!\!\!\!\!\!=\frac{ie^{2}m}{2k^3_{0}}\bar{n_e}c^{\mu\nu}v_\mu v_\nu\{-v.\epsilon^{1}v.\epsilon^{2}Q^{(1)}+2(v.\epsilon^{2}v.\epsilon^{2}-v.\epsilon^{1}v.\epsilon^{1})U^{(1)}\},
\end{eqnarray}
where
\beq
\int {\frac {d^3q}{(2\pi)^3} }n_{e}({\bf q})= \bar n_e \:\:\:\:;\:\:\:\: \int {\frac {d^3q}{(2\pi)^3} }q_in_{e}({\bf q})= mv_i\bar n_e ,
\eeq
and $v=(1,{\bf v})$.  Thus for $\lambda_0=1cm$, z=1000 and for length scale about 1 kpc and  number density of electrons about 0.1 per $cm^3$ one has

\beq
\Delta\phi_{FC}\sim 1\times10^{20}|c^{\mu\nu}|,
\eeq
or to have a phase of order one we should have $|c^{\mu\nu}|\sim 10^{-20}$.  One should note that if we take the intergalactic electron number density of about $10^{-8}$ per $cm^3$ then for a length scale about few Gpc the result only changes an order of magnitude.

%%%%%%%%%%%%%%%%%%%%%%%%%%%%%%%%%%%%%%%%%%%%%%%%%%%%%%%%%%%%%%%%%%%%%%%%%%%%%%%%%%%%%%%%%%%%

 \subsubsection{$d^{\mu\nu}$ term}

 We now ignore the $c^{\mu\nu}$ and move to the $d^{\mu\nu}$ term of the Lagrangian (\ref{cd}),
\begin{eqnarray} \mathcal{L} =
\frac{i}{2} \bar{\psi} \gamma^\mu \mathcal{D}_{\mu} \psi - m
\bar{\psi}\psi
+\frac{i}{2}\:d^{\mu\nu}\bar{\psi}\gamma^{5}\gamma_{\mu}\mathcal{D}_{\nu}\psi. \label{dmu}
\end{eqnarray}
Similar to the $c^{\mu\nu}$ case, one can evaluate the Compton amplitudes $\mathcal{M}_{1}$ and $\mathcal{M}_{2}$ and then the interaction Hamiltonian in order to find the damping term. The computations are detailed in appendix B. Using that results, the evolution of Stokes parameter $V$ is given by
 \begin{eqnarray}
    \label{LVD}
        \dot{V}^{(1)}=0.
\end{eqnarray}
Hence the $d_{\mu\nu}$ correction to the electron sector does not generate any circular component for
 the polarization of CMB. Although considering the higher order corrections of $d^{\mu\nu}$ may result in
 a non-zero source term for the V.
%%%%%%%%%%%%%%%%%%%%%%%%%%%%%%%%%%%%%%%%%%%%%%%%%%%%%%%%%%%%%%%%%%%%%

\section{Circular polarization due to noncommutativity}

In this section we consider the noncommutative corrections to the
QED interaction term. There is two fundamentally different
approaches in which one can find a closure form for the Lie algebra
of a noncommutative gauge theory \cite{nc1,nc2,nc3,nc4,nc5,nc6,nc7,nc8,nc9,nc10}. In the
approach proposed by Chaichian et al. \cite{nc1}, the group
closure property is given only for U(N) gauge theories with
matter content in the fundamental or adjoint representations. This
kind of model building has been criticized because it leads to
some conclusions for the standard model which is not consistent
with the our knowledge of particle physics. Wess and his
collaboration have developed another approach to construct a
consistent noncommutative gauge theory \cite{nc2}. In this
approach, the Lie algebra is extended to an enveloping algebra. The
mapping between the noncommutative gauge theory and the effective
theory on the usual commutative space-time is obtained by requiring
that the theory be invariant under both noncommutative and the
usual commutative gauge transformations. These requirements are
given through the solutions of some differential equations known
as the Seiberg-Witten map (SWM) \cite{nc2} appeared
originally in the context of string theory. The basic property of
the SWM method is that the new interactions are added but no
additional particles are introduced. Also in this approach both the
gauge field $A^{\mu}$ and the gauge transformation parameter
$\Lambda$ are expanded in terms of noncommutative parameter \bea
&& \hat{A}_{\mu}=A_{\mu}+ \frac{1}{4}\theta^{\alpha\beta}\{
\partial_\alpha A_\mu+F_{\alpha\mu},A_\beta
\}+\mathcal{O}(\theta^{2}),
 \no \\&& \hat{\Lambda}=\Lambda+\frac{1}{4}\theta^{\alpha\beta}\{\partial_{\alpha}\Lambda,A_{\beta}\}+\mathcal{O}(\theta^{2}).\eea
The advantage is that this construction can be applied to any
gauge theory with arbitrary matter representation.  The SWM
formulation of the noncommutative standard model thus uses the
standard model gauge group. Following \cite{nc2}, the noncommutative
generalization of QED sector is defined by the action \beq
\mathcal{S}=\int d^4 x \left[\bar{\hat{\psi}}\star
i(\mathcal{D}\!\!\!\!/\star\hat{\psi})-m\bar{\hat{\psi}}\star\hat{\psi}-\frac{1}{2}\texttt{Tr}
\hat{F}_{\mu\nu}\hat{F}^{\mu\nu}\right] ,\eeq where \beq
\hat{F}_{\mu\nu}=
F_{\mu\nu}+\frac{1}{2}\theta^{\alpha\beta}\{F_{\mu\alpha},F_{\nu\beta}\}-
\frac{1}{4}\theta^{\alpha\beta}\{A_\alpha,(\partial_{\beta}+\mathcal{D}_{\beta})F_{\mu\nu}\}
+\mathcal{O}(\theta^{2}),\eeq and \beq \hat{\psi} = \psi +
\frac{1}{2}e\theta^{\mu\nu}A_{\mu}\partial_{\nu}\psi+\mathcal{O}(\theta^{2}).\eeq
The Feynman rules for the fermion-photon-fermion vertex is
modified as what given in Fig. 1. The noncommutative parameter $\theta$ in SWM approach has been bounded through different experiments. Using the experiment data, the best bounds are of the orders of $1-10\:\:\texttt{TeV}$ \cite{nc11}.

\begin{figure}\hspace{-1cm}
\centerline{\epsfysize=1.5in\epsfxsize=1.5in\epsffile{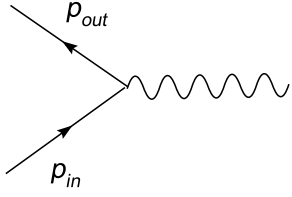}}
\caption{The fermion-photon-fermion vertex of the noncommutative QED:$\:\:ieQ_{f}\left\{\gamma_{\mu}-\frac{i}{2}\:\left[(p_{out}\theta p_{in})\gamma_{\mu}-(p_{out}\theta)_{\mu}(p_{in}\!\!\!\!\!\!\!/\:\:
  -m_{f})-(p_{in}\theta)_{\mu}(p_{out}\!\!\!\!\!\!\!\!\!/\:\:-m_{f})\right]\right\}$
 }\label{fig}
\end{figure}

Now we turn to the Boltzmann equation and perform the damping term. Using the new vertex,
the amplitude of the Compton scattering on a noncommutative
background is modified and gives corrections. Details of the calculation of the evolution of density matrix elements
are presented in Appendix C.
 Quoting the results (\ref{ro1}-\ref{ro2}) and similar to
the previous section, the evolution of the Stokes parameter $V$ is
given as
\bea
\dot{V}^{(1)}&&\!\!\!\!\!\!\!\!=i(\dot{\rho}_{12}-\dot{\rho}_{21})\no\\&&\!\!\!\!\!\!\!\!=\frac{ie^2}{2k_0m}\int
d\textbf{q}
\:n_e(\textbf{q})\Big(2(q\theta\ep_1q\cdot\ep_1+q\theta\ep_2q\cdot\ep_2)(\rho_{12}-\rho_{21})-
2(q\theta\ep_2q\cdot\ep_1+q\theta\ep_1q\cdot\ep_2)\rho_{11}\no\\&&
\:\:\:\:\:\:\:\:\:\:\:\:\:\:\:\:\:\:\:\:\:\:\:\:\:\:\:\:\:\:\:\:\:\:\:\:\:\:\:\:\:\:+2(q\theta\ep_2q\cdot\ep_1
+q\theta\ep_1q\cdot\ep_2)\rho_{22}\Big), \eea
where it can be more simplified as follows
 \bea \dot{V}^{(1)}=\frac{ie^2}{k_0m}\int d\textbf{q}
\:n_e(\textbf{q})(q\theta\ep_2q\cdot\ep_1+q\theta\ep_1q\cdot\ep_2)Q^{(1)}. \label{LVpolar}
 \eea

Hence similar to the Lorentz symmetry violating interaction terms
discussed in the previous section, the noncommutative quantum
electrodynamics based on the SWM,  also can generate the circular polarization for
the CMB radiation in the first order of noncommutative parameter $\theta$.  Now (\ref{LVpolar}) can be cast into
\bea \dot{V}^{(1)}=\frac{ie^2{\bar n_e}m}{\Lambda^2_{NC}k_0}(v.{\hat\theta}.\ep_2v\cdot\ep_1+v.{\hat\theta}.\ep_1v\cdot\ep_2)Q^{(1)},\label{LVpolar1}
 \eea
where $\Lambda_{NC}$ is the value of the non-commutativity parameter and $\hat\theta$ is unit vector that shows the non-commutativity direction and is a constant.  Therefore the Faraday phase conversion for $\Lambda_{NC}\sim 1-10\:\: TeV$
 will be around $10^{-7}-10^{-9}$ radians for $\lambda_0=1cm$, z=1000 and for length scale about 1 kpc.

%\section*{\small Acknowledgment}

%We would like to thank

%%%%%%%%%%%%%%%%%%%%%%%%%%%%%%%%%%%%%%%%%%%%%%%%%%%%%%%%%%%%%%%%%%%%%%%%%%%%%%%%%%
\section{ Circular polarization and magnetic field}

When photons impinge electrons in the
presence of a magnetic field, the circular polarization is generated naturally which leads to a non vanishing
Stokes parameter V whose magnitude depends upon the intensity of the radiation field \cite{Giovannini1,Giovannini2}. For instance,
 If the magnetic field is considered of the order of $B\sim \texttt{nG}$, the angular power spectra for small values of $l$ are derived as $\mathcal{O}(10^{-7})(\mu\texttt{K})^{2}$ and $\mathcal{O}(10^{-15})(\mu\texttt{K})^{2}$ for the $VT$ and $VV$ components respectively \cite{Giovannini1,Giovannini2}. In these references, the electron-photon scattering is computed by classical considerations. Here we want to investigate a similar phenomenon by considering the Compton scattering of electrons and photons.
A physical situation is considered when prior to recombination a cold plasma of electrons is supplemented by a weak magnetic field whose typical inhomogeneity
scale is at least comparable to Hubble radius $H^{-1}$ at the
corresponding epoch.
In the epoch of recombination, magnetic
field is estimated to be about $10^{-3}\mu\texttt{G}$ to
$10^{-1}\mu\texttt{G}$ \cite{reb}. This value of magnetic field is very weak
compared to the critical value
$B_{c}=m_{e}^{2}/e=4.414\times10^{13}\texttt{G}$. For this reason during our strategy,
the effect of magnetic field on the electron wave functions is ignored and only the electron
propagators are corrected. The propagator of a Dirac particle in an external constant
electromagnetic field background is well known from Schwinger work \cite{swn}
\begin{equation}\label{m1}
S(x,y)=\Phi(x,y)\int\frac{d^{4}p}{(2\pi)^{4}}e^{-ip.(x-y)}\tilde{S}(p),
\end{equation}
where the phase factor $\Phi(x,y)$ is defined by
\begin{equation}\label{m2}
\Phi(x,y)=\exp\left(-\frac{ie}{2}x^{\alpha}B_{\alpha\beta}y^{\beta}\right),
\end{equation}
in which $B_{\alpha}=-\frac{1}{2}B_{\alpha\beta}x^{\beta}$ and
 $B_{\alpha\beta}$ represents the strength background magnetic field tensor.
The other part of Schwinger propagator, $\tilde{S}(p)$, can be
expanded up to the linear background field as
\begin{equation}\label{m3}
\tilde{S}(p)=S_{0}(p)+S_{B}(p)+\mathcal{O}(B^{2}),
\end{equation}
for which $S_{0}$ is the usual electron propagator and $S_{B}$ is
the linear modified part of the Schwinger propagator
given by \cite{pal}
\begin{equation}\label{m4}
S_{B}(p)=-\frac{1}{2}eB_{\alpha\beta}
\frac{\gamma^{\alpha}(p\!\!\!/-m)\gamma^{\beta}}{(p^{2}-m^{2})^{2}}.
\end{equation}
Therefore invoking the above considerations, the Compton scattering matrix element in the presence of
background magnetic field is given by
\begin{equation}
\mathcal{M}=\mathcal{M}_{1}+\mathcal{M}_{2},
\end{equation}
where
\begin{equation}\label{m5}
\mathcal{M}_{1}=\bar{u}_{r'}(q')\int d^{4}xd^{4}x'
e^{ix'.q'}(-ie\epsilon\!\!\!/_{s'}(p')e^{ix'.p'})S(x',x)
\left(-ie\epsilon\!\!\!/_{s}(p)e^{-ix.p}\right)e^{-ix.q}u_{r}(q),
\end{equation}
and
\begin{equation}
\mathcal{M}_{2}=\bar{u}_{r'}(q')\int d^{4}xd^{4}x'
e^{ix'.q'}\left(-ie\epsilon\!\!\!/_{s}(p)e^{-ix'.p}\right)S(x',x)
\left(-ie\epsilon\!\!\!/_{s'}(p')e^{ix.p'}\right)e^{-ix.q}u_{r}(q),
\end{equation}
where $u_{r}(q)$ and $\bar{u}_{r'}(q')$ respectively are the wave functions
for the incoming electron with spin index $r$ and outgoing electron
with spin index $r'$. Also $\epsilon^{\mu}_{s}(p)$ and
$\epsilon^{\mu}_{s'}(p')$ are  polarization vectors for incoming
and outgoing photon chosen to be real and with indexes $s,s'=1,2$
labeling the physical transverse polarizations of photon. Using
the linear expanded Schwinger propagator (\ref{m3}) and by expanding the $\Phi(x,y)$ factor up to order $B$, for the amplitude (\ref{m5}) one can write
\begin{eqnarray}\label{m6}
\mathcal{M}_{1}&&\!\!\!\!\!\!\!\!\!=\bar{u}_{r'}(q')\int
d^{4}xd^{4}x'\frac{d^{4}k}{(2\pi)^{4}}
e^{ix'.(q'+p'-k)}e^{-ix.(q+p-k)}\left(-ie\epsilon\!\!\!/_{s'}(p')\right)\nonumber\\
&&\times\left(S_{0}(k)+S_{B}(k)-\frac{ie}{2}x'^{\alpha}B_{\alpha\beta}x^{\beta}S_{0}(k)\right)
\left(-ie\epsilon\!\!\!/_{s}(p)\right)u_{r}(q).
\end{eqnarray}
Integrating over the first two terms of (\ref{m6}) gives
\begin{equation}\label{m7}
(2\pi)^{4}\delta^{4}(q'+p'-q-p)\bar{u}_{r'}(q')\left(-ie\epsilon\!\!\!/_{s'}(p')\right)
\left(S_{0}(q+p)+S_{B}(q+p)\right)\left(-ie\epsilon\!\!\!/_{s}(p)\right)u_{r}(q).
\end{equation}
The remaining $\frac{ie}{2}x'^{\alpha}B_{\alpha\beta}x^{\beta}S_{0}(k)$ term explicitly violates the four momentum
conservation. One can assume that during the interaction time a small momentum $\delta q_{\mu}$ is transferred
from background into the internal electron of the associated Feynman diagram
so that the momentum conservation will be preserved and takes
the following form
\begin{equation}\label{m8}
q'_{\mu}+p'_{\mu}=q_{\mu}+p_{\mu}+\delta q_{\mu}.
\end{equation}
Substituting the condition (\ref{m8}) into the last term of
(\ref{m6}) and expand the exponential factor yields
\begin{equation}\label{m9}
\frac{ie}{2}B_{\alpha\beta}\int d^{4}xd^{4}x'
e^{i(x'-x).(q+p-k)}x'^{\alpha}x^{\beta}(1+ix'.\delta q).
\end{equation}
In the above expression, the term that is independent of $\delta
q$ is vanished because of antisymmetric properties of
$B_{\alpha\beta}$ with respect to $\alpha$ and $\beta$ while the term that is dependent on $\delta q$ remains. From classical electrodynamics we know that the four-momentum of electron that propagates between two space-time points,
changes similar to the Lorentz force $\Delta
p_{\mu}=-eB_{\mu\nu}\Delta x^{\nu}$. Then the
magnitude of $\delta q_{\mu}$ can be approximated to be of the order of external background
field $\delta q\sim eB$. Therefore this term becomes of the
order of $\mathcal{O}(B^{2})$ and hence can be ignored in the
scattering amplitude $\mathcal{M}_{1}$. With a similar procedure, the amplitude $\mathcal{M}_{2}$ is given by
\begin{equation}\label{m10}
\mathcal{M}_{2}=(2\pi)^{4}\delta^{4}(q'+p'-q-p)\bar{u}_{r'}(q')\left(-ie\epsilon\!\!\!/_{s}(p)\right)
\left(S_{0}(q-p')+S_{B}(q-p')\right)\left(-ie\epsilon\!\!\!/_{s'}(p')\right)u_{r}(q).
\end{equation}
The $B$-independent terms of $\mathcal{M}_{1}$ and $\mathcal{M}_{2}$ give the usual
Compton scattering. The correction up to the linear background field is represented as $\mathcal{M}_{B}$ and is written as
\begin{eqnarray}\label{m11}
\mathcal{M}_{B}&&\!\!\!\!\!\!\!\!\!=\mathcal{M}_{1B}+\mathcal{M}_{2B}\nonumber\\&&
\!\!\!\!\!\!\!\!\!=(2\pi)^{4}\delta^{4}(q'+p'-q-p)\bar{u}(q')\Big[
\left(-ie\epsilon\!\!\!/_{s'}(p')\right)S_{B}(q+p)
\left(-ie\epsilon\!\!\!/_{s}(p)\right)\nonumber\\&&\:\:\:\:\:\:\:\:
+\left(-ie\epsilon\!\!\!/_{s}(p)\right)
S_{B}(q-p')\left(-ie\epsilon\!\!\!/_{s'}(p')\right)\Big]u_{r}(q).
\end{eqnarray}
Using these amplitudes, the damping term on the right hand side of the Boltzmann equation can be computed. Computing the damping term is detailed in appendix D.
By using these results the density matrix element and then
the evolution of the Stokes parameter $V$ up to order of $e^{4}$ is derived as
\begin{eqnarray}\label{m21}
\dot{V}^{(1)}\!\!\!\!\!\!\!\!\!\!&&=i(\dot{\rho}_{12}-\dot{\rho}_{21})\nonumber\\
&&=\frac{\pi e^{4}}{4m^{2}k}\int
d\textbf{q}d\textbf{p}\delta(k-p)\left(\frac{1}{q.k}-\frac{1}{q.p}\right)
\left(\frac{1}{(q.k)^{2}}-\frac{1}{(q.p)^{2}}\right)\left(\tilde{q}.\epsilon_{1}(k)q.\epsilon_{2}(k)
-\tilde{q}.\epsilon_{2}(k)q.\epsilon_{1}(k)\right)\nonumber\\
&&\times n_{e}(\textbf{q})\Big[((q.\epsilon_{1}(p))^{2}+(q.\epsilon_{2}(p))^{2})(I^{(1)}(\textbf{k})-I^{(1)}(\textbf{p}))
-(q.\epsilon_{1}(p)q.\epsilon_{1}(p)-q.\epsilon_{2}(p)q.\epsilon_{2}(p))Q^{(1)}(\textbf{p})\nonumber\\
&&\:\:\:\:\:\:\:\:\:\:\:\:\:\:\:\:\:\:-\:2q.\epsilon_{1}(p)q.\epsilon_{2}(p)U^{(1)}(\textbf{p})\Big]+\mathcal{O}(k,p),
\end{eqnarray}
where $\tilde{q}_{\mu}=-eB_{\mu\nu} q^{\nu} $.  Hence similar to the LIV interaction terms
and the noncommutativity, corrections due to the external
 magnetic field can generate the circular
polarization for the CMB radiation as well.  But in contrast,  it is not needed a prior linear polarization to have a nonzero Stokes-V parameter that is in agreement with references \cite{Giovannini1,Giovannini2}.  But it should be noted that in \cite{Giovannini1,Giovannini2} the scattering matrix is classically obtained in the electron-ion magnetized plasma, while in our model the Compton scattering is calculated by considering the correction on the electron propagator in an external magnetic field.
  Here one can easily see that the circular polarization is linearly proportional to the background magnetic field.  Meanwhile if we integrate (\ref{m21}) over the momentum $p$ and use the approximation $q_0\simeq m \gg |\bf q| , |\bf k|$ then one can see that the correction will be proportional to $\lambda^3$  as follows

\begin{equation}
\dot{V}^{(1)}=\frac{e^4 m\lambda^3}{8\pi}\int \frac{d\Omega}{4\pi}{\bar n_e}({\bf v}.({\bf \hat{k}- \hat{p}}))^2 (\frac {eB}{m^2})F(v, \hat{\bf p}, \epsilon_{1}, \epsilon_{2}, I, U, Q, \frac {B_{\mu\nu}}{B}),\label{m22}
\end{equation}
where $F$ can be easily define by comparing (\ref{m21}) and (\ref{m22}).  Here again one can estimate the Faraday phase conversion for $z=1000$ as
\beq
\Delta\phi_{FC}\sim 1 \:\:rad\:\:\: (\frac{B}{10\mu G})\:\:( \frac {\lambda_0}{1cm})\:\:(\frac{\bar n_e}{0.1 cm^3})\:\:(\frac{L}{1kpc}).
\eeq

%%%%%%%%%%%%%%%%%%%%%%%%%%%%%%%%%%%%%%%%%%%%%%%%%%%%%%%%%%%%%%%%%%%%%%%%%
\section{\large Discussion and Conclusion }
In this work we have considered the evolution of the Stokes
parameters in the context of a general extension of quantum electrodynamics allowing for Lorentz violation (LVQED), NCQED and in the presence of the background magnetic field.  In fact all nonzero backgrounds can potentially produce the circular polarization for the CMB radiation.  In the LVQED we have explored three cases, one case in the photon sector and two cases in the electron sector.  In the former one the induced circular polarization is provided through the interaction of photon with the background Lorentz violation field, Eq.(\ref{LVP}), while in the latter cases the corrections on the Compton scattering in the LVQED cause the polarization, see  Eqs.(\ref{eq:LVC}) and (\ref{LVD}). It is shown that the Stokes parameter V in both sectors are linearly proportional to the dimensionless Lorentz parameters $k_F$ and $c$ that is in contrast with the result of \cite{kosowsky} where the second term of Lagrangian (\ref{ext}) is considered as the interaction term and it is found that V depends quadratically on the $k_{AF}$ that is too small with respect to our result.  We have also estimated the Faraday phase conversion in the LV case to give an upper bound on the dimensionless Lorentz parameters $c $ and $k_F$ about $10^{-20}$ that is comparable to the current bounds on these parameters.  In the NCQED we have found the nonzero parameter V through a correction on the Compton scattering in the NC space that is linearly proportional with the parameter of non commutativity as given in (\ref{LVpolar1}).  Our estimation on the Faraday phase conversion is about $10^{-7}-10^{-9}$ radians for $\Lambda_{NC}\sim 1-10 TeV$ that is too small to be detectable.  Finally we have examined the Stokes parameter V for the background magnetic field.  In this case the refractive term is zero but the damping term has a nonzero V parameter that depends linearly  on the magnetic field, see Eq.(\ref{m22}).  This result can be compared with the result of \cite{Cooray:2002nm} in which the circular polarization of CMB due to the Faraday conversion depends quadratically on the background magnetic field.  Furthermore,  in contrast with \cite{Cooray:2002nm}, all the other Stokes parameters (I, Q and U) have some contributions on the generation of the Stokes-V parameter in the background magnetic field.  In fact it is not necessary a prior linear polarization to have a nonzero Stokes-V parameter that is in agreement with references \cite{Giovannini1,Giovannini2}.  By comparing our results together we see a  $\lambda^3$-dependence for the magnetic field and LV on one hand and $\lambda$-dependence for non-commutativity on the other hand  which leads to different spectrum for different wavelengths.
Because the required sensitivity level to detect circular
polarization of CMB is beyond the current instrumental
techniques, the experimental bound on the $V$ modes is rare. The bound reported in \cite{natureexp}, constrains it as $V\leq 1\:\:\mu\texttt{K}$.
However, the post-Planck CMB polarization
experiments are already under study \cite{cmbpol}. Then it is time
to address the question whether CMB
is circularly polarized or not and what physical information can
be extracted from its measurement. Also
it seems that
 if a new experimental device can be built with sufficient polarimetric sensitivity beyond the $\mu\texttt{K}$ range and with good
control of systematic errors, the resulting observations may have the potential to realize the
physics at energy scales far beyond the reach of the largest feasible particle
accelerators and provide a window to new physics at these energy scales.
Therefore we encourage experimenters to make experiments in order to measure the circular polarization of cosmic microwave radiation.
%%%%%%%%%%%%%%%%%%%%%%%%%%%%%%%%%%%%%%%%%%%%%%%%%%%%%%%%%%%%%%%%%%%%%%%%%%%%%%%%%%
\appendix
\section{The calculation of density matrix elements with $c^{\mu\nu}$ coefficients}
In this appendix at first
the scattering amplitudes ${\cal {M}}_1$ and ${\cal {M}}_2$ are calculated as follows.
For the Lagrangian (\ref{cd}) with $c^{\mu\nu}$ term,
the equation of motion for the electron $\psi$ is found as
\beq
(i\!\!\not\!\partial-m
+ i\:c^{\nu\mu}\gamma_{\nu}\partial_{\mu})\psi=0 .\eeq
for which the modified Dirac equation is written as
\beq
(\:\!\!\not\!q-m
+ \:c^{\nu\mu}\gamma_{\nu}q_{\mu})\:u_{s}(q)=0,\label{dirac} \eeq
in which $ \psi=\:u_{s}(q)\exp(-iq.x)$.
In the Weyl chiral representation for the Dirac matrices, after some straightforward calculations, one finds the following solution for the Dirac equation (\ref{dirac})
\begin{equation}
\label{eq-gamma} u_{s}(q)=\left(
\begin{array}{cc}
\sqrt{q.\sigma+c^{\nu\mu}q_{\mu}\sigma_{\nu}}\: \zeta_{s} \\
\sqrt{q.\bar{\sigma}+c^{\nu\mu}q_{\mu}\bar{\sigma_{\nu}}}
\:\zeta_{s}
\end{array}\right),
\end{equation}
where $\sigma=(1,\vec{\sigma})$ and $\bar{\sigma}=(1,-\vec{\sigma})$.
The electron propagator can be  also calculated straightforwardly with a correction in the denominator as
\begin{equation}
S(p)=\frac{i}{ q\!\!\!/-m\,+c^{\nu\mu}q_{\mu}\gamma_{\nu}}.
\end{equation}
Since $|c^{\nu\mu}|\ll\:1$, the propagator
 can be expanded up to the linear order of $c^{\nu\mu}$
as
\begin{eqnarray}
\nonumber
 S(q)&&\!\!\!\!\!\!\!\!=S_{0}(q)+S_{0}(q)(ic^{\nu\mu}q_{\mu}\gamma_{\nu})S_{0}(q)+\cdot\cdot\cdot\no \\&&
\!\!\!\!\!\!\!\!=\frac{i}{q\!\!\!/-m}+\frac{i}{q\!\!\!/-m}(ic^{\nu\mu}q_{\mu}\gamma_{\nu})\frac{i}{q\!\!\!/-m}+\cdot\cdot\cdot,
\end{eqnarray}
where $S_{0}(q)$ is the usual electron propagator. The $u_s(q)$ solutions can be also expanded in terms of $c^{\nu\mu}$,
\beq u_s(q)= u_0(q)+u(q)+\cdot\cdot\cdot , \eeq
where $u_0(q)\equiv u_{s0}(q)$ is the free electron solution for the usual Dirac equation in the absence of $c^{\nu\mu}$ term and
$u(q)\equiv u_{s}(q)$ is the correction of the $c^{\nu\mu}$ order. The dots show the higher order corrections.
Consequently, the Compton scattering amplitude ${\cal M}$ is modified with corrections
given by
\begin{eqnarray}
{\cal M} &&\!\!\!\!\!\!\!\!=
\mathcal{M}_{1}+\mathcal{M}_{2}\no\\&&\!\!\!\!\!\!\!\!=
\bar{u}(q')(-ie)\left(\gamma^{\mu}+c^{\nu\mu}\gamma_{\nu}\right)\epsilon^{*s'}
_{\mu}(p')S(q+p)(-ie)\left(\gamma^{\rho}+c^{\nu\rho}\gamma_{\nu}\mathbf{}\right)\epsilon_{\rho}^{s}(p)u(q)
%\label{eq-iM} & &
\no\\&&\!\!\!\!\!
+\,\bar{u}(q')(-ie)\left(\gamma^{\rho}+c^{\nu\rho}\gamma_{\nu}\right)\epsilon
_{\rho}^{s}(p)S(q-p')(-ie)\left(\gamma^{\mu}+c^{\nu\mu}\gamma_{\nu}\right)\epsilon^{*s'}_{\mu}(p')u(q),
\end{eqnarray}
where $\epsilon(p)$ and $\epsilon'(p')$ are transverse polarization
vectors for the external real photons. The scattering amplitudes up to the order of $c^{\nu\mu}$ are simplified as follows
\bea \mathcal{M}_{1}&&\!\!\!\!\!\!\equiv -e^{2}
\bar{u_{0}}(q')\{\gamma^{\mu}\epsilon^{*s'}
_{\mu}(p')S_{0}(q+p)c^{\nu\rho}\gamma_{\nu}\epsilon_{\rho}^{s}+
 \gamma^{\mu}\epsilon^{*s'}
_{\mu}(p')S_{0}(q+p)\left(ic^{\nu\beta}q_{\beta}\gamma_{\nu}
\right)\times\nonumber\\&&S_{0}(q+p)\gamma^{\rho}\epsilon_{\rho}^{s}(p)
c^{\nu\mu}\gamma_{\nu}\epsilon^{*s'}_{\mu}(p')S_{0}(q+p)\gamma^{\rho}\epsilon_{\rho}^{s}(p)\}u_{0}(q)-\nonumber\\&&
\:\:\:\:\:\:\:\:
e^2\bar{u}(q')\gamma^{\mu}\epsilon^{*}
_{\mu}(p')S_{0}(q+p)\gamma^{\rho}\epsilon_{\rho}^{s}u(q), \eea
 and
\bea \mathcal{M}_{2}&&\!\!\!\!\equiv -e^{2}
\bar{u_{0}}(q')\{\gamma^{\rho}\epsilon^{s}
_{\rho}(p)S_{0}(q-p')c^{\nu\mu}\gamma_{\nu}\mathbf{}\epsilon_{\mu}^{*s'}(p')
 \gamma^{\rho}\epsilon^{s}
_{\rho}(p')S_{0}(q-p')\left(ic^{\nu\beta}q_{\beta}\gamma_{\nu}\right)\times\nonumber\\&&S_{0}(q-p')
\gamma^{\mu}\epsilon_{\mu}^{*s'}(p')+
c^{\rho\nu}\gamma_{\nu}
\epsilon^{s}_{\rho}(p)S_{0}(q-p')\gamma^{\mu}\epsilon_{\mu}^{*s'}(p')\}u_{0}(q)-\nonumber\\&& \:\:\:\:\:\:e^2\bar{u}(q')
\gamma^{\rho}\epsilon^{s}
_{\rho}(p)S_{0}(q-p')\gamma^{\mu}\epsilon_{\mu}^{*s'}u(q). \eea
Substituting these amplitudes in the Eq.~(\ref{eqcomm}) and using
the expectation values computed in \cite{Kosowsky:1994cy} and using the photon polarization vector properties $\epsilon_i(k)\cdot k=0$ and $\epsilon_i\cdot\epsilon_j =\delta_{ij}$, it follows that
\begin{eqnarray}
    \label{eq:4.42}
\left\langle\left[ \mathcal{\hat
H}_{\text{int}}(t),\mathcal{\hat D}_{ij}({\bf
k})\right]\right\rangle&&\!\!\!\!\!\!\!=e^{2}(2\pi)^{3}\delta^{3}(0)\int d {\bf
q}n_{e}(q)\left[ \delta_{si}\rho_{s'j}( {\bf
k})-\delta_{s'j}\rho_{si}({\bf k})
\right]\no\\&&\!\!\!\!\!\!\!\!\!\!\!\!\!\!\!\!\!\!\!\!\!\!\!\!\!\!\!\!\!\!\!\!\!\!\!\!\!\!\!\!
\times\Big((\frac{1}{2k.q})\{c^{\rho\nu}\epsilon_{\rho}^{s}(k)
(q.k\epsilon_{\nu}^{s'}(k)-q.\epsilon^{*s'}k_{\nu})+c^{\nu\mu}\epsilon_{\mu}^{*s'}(k)
(q.k\epsilon_{\nu}^{s}(k)-q.\epsilon^{s}k_{\nu})+\no\\&&\!\!\!\!\!\!\!\!\!\!\!\!\!\!\!\!\!\!\!\!\!\!\!\!\!\!\!\!\!\!\!\!\!\!\!\!\!\!\!\!
c^{\nu\beta}q_{\beta}(q.\epsilon^{s}
\epsilon^{*s'}_{\nu}+q.\epsilon^{*s'}\epsilon^{s}_{\nu})\}+\frac{2c^{\nu\beta}}{(q.k)^{2}}
(q_{\beta}q_{\nu}+k_{\beta}k_{\nu})(q.\epsilon^{s}q.\epsilon^{*s'})\Big),
\end{eqnarray}
where the quantity $n_e({\bf q})$ is the number density of electrons of
momentum ${\bf q}$ per unit volume. It is now straightforward to calculate the time derivative of the density matrix which result in
\bea
    \label{eq:}
        \dot{\rho}_{11}({\bf k})&&\!\!\!\!\!\!\!\!= \frac{e^{2}}{2mk^{0}}\int {d{\bf q}}\:n_{e}({\bf q})\:\Big[\:
\frac{c^{\mu\nu}}{q.k}[\{(q.k\epsilon_{\mu}^{1}\epsilon_{\nu}^{2}-k_{\nu}q.\epsilon^{2}\epsilon_{\mu}^{1})
+(q.k\epsilon_{\nu}^{1}\epsilon_{\mu}^{2}-k_{\nu}q.\epsilon^{1}\epsilon_{\mu}^{2})\:\no\\&&+
q_{\nu}(q.\epsilon^{1}\epsilon_{\mu}^{2}+q.\epsilon^{2}\epsilon_{\mu}^{1})\}\rho_{21}
-\{(q.k\epsilon_{\mu}^{2}\epsilon_{\nu}^{1}-k_{\nu}q.\epsilon^{1}\epsilon_{\mu}^{2})
+(q.k\epsilon_{\mu}^{1}\epsilon_{\nu}^{2}-k_{\nu}q.\epsilon^{2}\epsilon_{\mu}^{1})\no\\&&+
        q_{\nu}(q.\epsilon^{2}\epsilon_{\mu}^{1}+q.\epsilon^{1}\epsilon_{\mu}^{2})\}\rho_{12}]
                +\frac{2c^{\mu\nu}}{(q.k)^{2}}[(q_{\mu}q_{\nu}+k_{\mu}k_{\nu})q.\epsilon^{1}
        q.\epsilon^{2}(\rho_{21}-\rho_{12})]\Big],
\eea

\begin{eqnarray}
    \label{eq:}
        \dot{\rho}_{22}({\bf k})&&\!\!\!\!\!\!\!\!= \frac{e^{2}}{2mk^{0}}\int {d{\bf
        q}}n_{e}({\bf
        q})\:\Big[\:\frac{c^{\mu\nu}}{q.k}[\{(q.k\epsilon_{\mu}^{2}\epsilon_{\nu}^{1}-k_{\nu}
        q.\epsilon^{1}\epsilon_{\mu}^{2})
        +(q.k\epsilon_{\nu}^{2}\epsilon_{\mu}^{1}-k_{\nu}q.\epsilon^{2}\epsilon_{\mu}^{1})\no\\&&
   \:\:\:\:\:\:\:\:\:\:\:\:\:\: +q_{\nu}(q.\epsilon^{2}\epsilon_{\mu}^{1}+q.\epsilon^{1}
        \epsilon_{\mu}^{2})\}\rho_{12}-\{(q.k\epsilon_{\mu}^{1}\epsilon_{\nu}^{2}-
        k_{\nu}q.\epsilon^{2}\epsilon_{\mu}^{1})\no\\&&
        \:\:\:\:\:\:\:\:\:\:\:\:\:\:+(q.k\epsilon_{\mu}^{2}\epsilon_{\nu}^{1}-
        k_{\nu}q.\epsilon^{1}\epsilon_{\mu}^{2})+q_{\nu}(q.\epsilon^{1}\epsilon_{\mu}^{2}+
        q.\epsilon^{2}\epsilon_{\mu}^{1})\}\rho_{21}]\no\\&&
               \:\:\:\:\:\:\:\:\:\:\:\:\:\: +\frac{2c^{\mu\nu}}{(q.k)^{2}}\{(q_{\mu}q_{\nu}+k_{\mu}k_{\nu})q.\epsilon^{1}q.\epsilon^{2}
        (\rho_{12}-\rho_{21})\}\Big],
\end{eqnarray}

\begin{eqnarray}
    \label{eq:}
        \dot{\rho}_{12}({\bf k})&&\!\!\!\!\!\!\!\!= \frac{e^{2}}{2mk^{0}}\int {d{\bf
        q}}n_{e}\:\Big[\:({\bf
        q})\frac{c^{\mu\nu}}{q.k}[\{(q.k\epsilon_{\mu}^{1}\epsilon_{\nu}^{2}-k_{\nu}q.\epsilon^{2}
        \epsilon_{\mu}^{1})
        +(q.k\epsilon_{\nu}^{1}\epsilon_{\mu}^{2}-k_{\nu}q.\epsilon^{1}
        \epsilon_{\mu}^{2})\no\\&&+q_{\nu}(q.\epsilon^{1}\epsilon_{\mu}^{2}
        +q.\epsilon^{2}\epsilon_{\mu}^{1})\}(\rho_{22}-\rho_{11})+\{q.k(\epsilon_{\mu}^{1}\epsilon_{\nu}^{1}
        -\epsilon_{\mu}^{2}\epsilon_{\nu}^{2})-k_{\nu}q.\epsilon^{1}\epsilon_{\mu}^{1}
        +k_{\nu}q.\epsilon^{2}\epsilon_{\mu}^{2}\no\\&&-\epsilon_{\mu}^{2}
        \epsilon_{\nu}^{2})-(k_{\nu}q.\epsilon^{1}\epsilon_{\mu}^{1}-k_{\nu}q.\epsilon^{2}\epsilon_{\mu}^{2})+2q_{\nu}(q.\epsilon^{1}\epsilon_{\mu}^{1}-q.\epsilon^{2}
        \epsilon_{\mu}^{2})\}
        \rho_{12}]\no\\&&
        +\frac{2c^{\mu\nu}}{(q.k)^{2}}\{(q_{\mu}q_{\nu}+k_{\mu}k_{\nu})[q.\epsilon^{1}q.\epsilon^{2}
        (\rho_{22}-\rho_{11})+((q.\epsilon^{1})^{2}-(q.\epsilon^{2})^{2})\rho_{12}]\}\Big],
\end{eqnarray}
and
\begin{eqnarray}
    \label{eq:}
        \dot{\rho}_{21}({\bf k})&&\!\!\!\!\!\!\!\!= \frac{e^{2}}{2mk^{0}}\int {d{\bf
        q}}n_{e}\:\Big[\:({\bf
        q})\frac{c^{\mu\nu}}{q.k}[\{(q.k\epsilon_{\mu}^{2}\epsilon_{\nu}^{1}-k_{\nu}q.\epsilon^{1}
        \epsilon_{\mu}^{2})\no\\&& +(q.k\epsilon_{\nu}^{2}\epsilon_{\mu}^{1}-k_{\nu}q.\epsilon^{2}
        \epsilon_{\mu}^{1})+ q_{\mu}(q.\epsilon^{2}\epsilon_{\nu}^{1}
        +q.\epsilon^{1}\epsilon_{\nu}^{2})\}(\rho_{11}-\rho_{22})\no\\&&+\{2q.k(\epsilon_{\rho}^{2}\epsilon_{\nu}^{2}-
        \epsilon_{\mu}^{1}\epsilon_{\nu}^{1})-2(k_{\nu}q.\epsilon^{2}
        \epsilon_{\mu}^{2}-k_{\nu}q.\epsilon^{1}\epsilon_{\mu}^{1})+2q_{\nu}
        (q.\epsilon^{2}\epsilon_{\mu}^{2}-q.\epsilon^{1}\epsilon_{\mu}^{1}) \}\rho_{21}]\no\\&&
      +\frac{2c^{\mu\nu}}{(q.k)^{2}}(q_{\mu}q_{\nu}+k_{\mu}k_{\nu})[q.\epsilon^{1}q.\epsilon^{2}
        (\rho_{11}-\rho_{22})+((q.\epsilon^{2})^{2}-(q.\epsilon^{2})^{2})\rho_{21}]\Big].
\end{eqnarray}

%%%%%%%%%%%%%%%%%%%%%%%%%%%%%%%%%%%%%%%%%%%%%%%%%%%%%%%%%%%%%%%%%%%%%%%%%%%%%%%%%%
\section{The calculation of density matrix elements with $d^{\mu\nu}$ coefficients}
In this appendix we compute the density matrix elements with interaction given by (\ref{dmu}).
The equation of motion for the Lagrangian (\ref{dmu}) is calculated as
\begin{equation}
\label{eq-gamma} u_{s}(q)=\left(
\begin{array}{cc}
\sqrt{q.\sigma+(d^{\nu\mu}q_{\mu}\sigma_{\nu})}\: \zeta_{s} \\
\sqrt{q.\bar{\sigma}-(d^{\nu\mu}q_{\mu}\bar{\sigma}_{\nu})}
\:\zeta_{s}
\end{array}\right),
\end{equation}
and the propagator changes as
\begin{equation} S(q)=\frac{i}{
p\!\!\!/-m\,+d^{\nu\mu}p_{\mu}\gamma^5\gamma_{\nu}}.
\end{equation}
similar to the $c^{\mu\nu}$ case, one can write the following expansion for the propagator up to the order $\mathcal{O}(d^{\mu\nu})$
\begin{eqnarray}
\nonumber
 S(q)&&\!\!\!\!=S_{0}(q)+S_{0}(p)(id^{\nu\mu}q_{\mu}\gamma^5\gamma_{\nu})S_{0}(q)\no\\&&
 \equiv\frac{i}{q\!\!\!/-m}+\frac{i}{
q\!\!\!/-m}(id^{\nu\mu}q_{\mu}\gamma^5\gamma_{\nu})\frac{i}{ q\!\!\!/-m}.
\end{eqnarray}
 Then the Compton scattering matrix elements ${\cal M}_{1}$ and ${\cal M}_{2}$ in this theory
are calculated as
\begin{eqnarray}
{\cal M} &&\!\!\!\!\!\!\!\!\!=
\mathcal{M}_{1}+\mathcal{M}_{2}\no\\&&\!\!\!\!\!\!\!\!
=\bar{u}(q')(-ie)\left(\gamma^{\mu}+d^{\nu\mu}\gamma^5\gamma_{\nu}\right)\epsilon^{*s'}
_{\mu}(k')S(q+p)(-ie)\left(\gamma^{\rho}+d^{\nu\rho}\gamma^5\gamma_{\nu}\mathbf{}\right)\epsilon_{\rho}^{s}(p)u(q)
\no\\ \label{eq-iM} & &\!\!\!\!\!\!\!\!\!\!\!\!\
+\,\bar{u}(q')(-ie)\left(\gamma^{\rho}+d^{\nu\rho}\gamma^5\gamma_{\nu}\right)\epsilon
_{\rho}^{s}(k)S(q-p')(-ie)\left(\gamma^{\mu}+d^{\nu\mu}\gamma^5\gamma_{\nu}\right)\epsilon^{*s'}_{\mu}(pk')u(q),
\end{eqnarray}
where
\bea \mathcal{M}_{1}&&\!\!\!\!\!\!\!\!\!\equiv(-e^{2})
\bar{u}_{0}(q')\{\gamma^{\mu}\epsilon^{*s'}
_{\mu}(p')S_{0}(q+p)\left(d^{\nu\rho}\gamma^{5}\gamma_{\nu}\mathbf{}\right)\epsilon_{\rho}^{s}\nonumber\\&&+
 \gamma^{\mu}\epsilon^{*s'}
_{\mu}(k')S_{0}(q+p)\left(id^{\nu\beta}q_{\beta}\gamma^{5}\gamma_{\nu}\mathbf{}
\right)S_{0}(q+p)\gamma^{\rho}\epsilon_{\rho}^{s}(p)\nonumber\\&&+
d^{\nu\mu}\gamma^{5}\gamma_{\nu}\epsilon^{*s'}_{\mu}(p')S_{0}(q+p)\gamma^{\rho}
\epsilon_{\rho}^{s}(p)\}u_{0}(q)\nonumber\\&&
-e^2\bar{u}(q')\gamma^{\mu}\epsilon^{*}
_{\mu}(p')S_{0}(q+p)\gamma^{\rho}\epsilon_{\rho}^{s}u(q), \eea
and
\bea \mathcal{M}_{2}&&\!\!\!\!\equiv(-e^{2})
\bar{u}_{0}(q')\{\gamma^{\rho}\epsilon^{s}
_{\rho}(p)S_{0}(q-p')\left(d^{\nu\mu}\gamma^{5}\gamma_{\nu}\mathbf{}\right)\epsilon_{\mu}^{*s'}(p')+\nonumber\\&&
 \gamma^{\rho}\epsilon^{s}
_{\rho}(p')S_{0}(q-p')\left(id^{\nu\beta}q_{\beta}\gamma^{5}\gamma_{\nu}\mathbf{}\right)S_{0}(q-p')
\gamma^{\mu}\epsilon_{\mu}^{*s'}(p')+\nonumber\\&&
d^{\nu\rho}\gamma^{5}\gamma_{\nu}
\epsilon^{s}_{\rho}(p)S_{0}(q-p')\gamma^{\mu}\epsilon_{\mu}^{*s'}(p')\}u_{0}(q)+\nonumber\\&&(-e^2)\bar{u}(q')
\gamma^{\rho}\epsilon^{s}_{\rho}(p)S_{0}(q-p')\gamma^{\mu}\epsilon_{\mu}^{*s'}u(q).
\eea
Substituting these amplitudes in the Eq.~(\ref{eqcomm}), and using the expectation values calculated in \cite{Kosowsky:1994cy}, leads to
\begin{eqnarray}
    \label{eq:4.42}
\left\langle\left[ \mathcal{\hat
H}_{\text{int}}(t),\mathcal{\hat D}_{ij}({\bf
k})\right]\right\rangle &&\!\!\!\!\!\!\!= \frac{e^{2}}{2\:m}(2\pi)^3{\delta}^3(0)\int d {\bf
q}n_{e}(q)\left[
\delta_{is}\rho_{s'j}( {\bf k})-\delta_{js'}\rho_{is}({\bf k})
\right]\no\\&&\!\!\!\!\!\!\!\!\!\!\!\!\!\!\!\!\!\!\!\!\!\!\!\!\!\!\!\!\!\!\!\!\!
\Big\{\frac{d^{\nu\mu}}{(k.q)^2}(\epsilon^{*s'}_{\beta}\epsilon^{s}_{\sigma}-
\epsilon^{*s'}_{\sigma}\epsilon^{s}_{\beta})[(-2{q}_{\alpha}{k}_{\rho}({q}_{\mu}{q}_{\nu}+
{k}_{\mu}{k}_{\nu})-2{q}_{\alpha}{q}_{\rho}({q}_{\mu}{k}_{\nu}+{k}_{\mu}{q}{\nu}))\no\\&&
\!\!\!\!\!\!\!\!\!\!\!\!\!\!\!\!\!\!\!\!\!\!\!\!\!\!\!\!\!\!\!\!\!
{\epsilon}^{\alpha\beta\rho\sigma}+(k^2{q}_{\alpha}{k}_{\mu}-2m^2{q}_{\mu}{k}_{\alpha}+
2(q.k)q_{\alpha}q_{\mu}){{\epsilon^{\alpha\beta}}_{\nu}}^{\rho}]+\frac{2d^{\nu\mu}}{q.k}
{\epsilon_{\nu}}^{\alpha\beta\rho}\no\\&&
\!\!\!\!\!\!\!\!\!\!\!\!\!\!\!\!\!\!\!\!\!\!\!\!\!\!\!\!\!\!\!\!\![\epsilon^{*s'}_{\mu}q_{\alpha}
(q_{\beta}\epsilon^{s}_{\rho}-q_{\rho}\epsilon^{s}_{\beta})-\epsilon^{s}_{\mu}q_{\alpha}
(q_{\beta}\epsilon^{*s'}_{\rho}-q_{\rho}\epsilon^{*s'}_{\beta})]-\frac{4id^{\nu\mu}}
{q.k}{\epsilon^{\alpha\beta\rho}}_{\nu}q_{\mu}q_{\beta}\epsilon^{*s'}_{\alpha}
\epsilon^{s}_{\rho}\Big\},
\end{eqnarray}
where $\epsilon^{\alpha\beta\sigma\rho}$  is the Levi-Civita symbol appeared through tracing process when $\gamma^{5}$ is present.
Similar to the $c_{\mu\nu}$ case, the time evolution of the $\rho_{12}$ and $\rho_{21}$ are obtained as
\bea
    \label{eq:}
        \dot{\rho}_{12}({\bf k})\!\!\!\!\!\!\!&&= \frac{e^{2}}{2mk^{0}}\int {d{\bf
        q}}n_{e}({\bf
        q})\Big\{{\rho}_{12}\Big[\frac{4id^{\nu\mu}}{q.k}{\epsilon^{\alpha\beta\rho}}_{\nu}
        q_{\mu}q_{\beta}({\epsilon}^1_{\alpha}{\epsilon}^1_{\rho}-
        {\epsilon}^2_{\alpha}{\epsilon}^2_{\rho})]+(\rho_{22}-\rho_{11})\no\\&& \!\!\!\!\!\!\!\!\!\!\!\!
        [\frac{d^{\nu\mu}}{(q.k)^2}({\epsilon}^2_{\beta}{\epsilon}^1
        _{\sigma}-{\epsilon}^2_{\sigma}{\epsilon}^1_{\beta})
        [(-2{q}_{\alpha}{k}_{\rho}({q}_{\mu}{q}_{\nu}
        +{k}_{\mu}{k}_{\nu})-2{q}_{\alpha}{q}_{\rho}({q}_{\mu}{k}_{\nu}+{k}_{\mu}{q}{\nu}))
        {\epsilon}^{\alpha\beta\rho\sigma}\no\\&& \!\!\!\!\!\!\!\!\!\!\!\!+(k^2{q}_{\alpha}{k}_{\mu}-2m^2{q}_
        {\mu}{k}_{\alpha}+2(q.k)q_{\alpha}q_{\mu}){{\epsilon^{\alpha\beta}}_{\nu}}^{\rho}]+
        \frac{2d^{\nu\mu}}{q.k}{\epsilon^{\alpha\beta\rho}}_{\nu}\no\\&& \!\!\!\!\!\!\!\!\!\!\!\!
        [q_{\alpha}{q}_{\rho}({\epsilon}^2_{\mu}{\epsilon}^1_{\beta}-{\epsilon}^1_{\mu}
        {\epsilon}^2_{\beta})+q_{\alpha}q_{\beta}({\epsilon}^1_{\mu}{\epsilon}^2_{\rho}-
        {\epsilon}^2_{\mu}{\epsilon}^1_{\rho})+2iq_{\mu}q_{\beta}{\epsilon}^2_{\alpha}
        {\epsilon}^1_{\rho}]\Big]\Big\},
                \eea
and
\bea
    \label{eq:}
        \dot{\rho}_{21}({\bf k})\!\!\!\!\!\!\!&&= \frac{e^{2}}{2mk^{0}}\int {d{\bf
        q}}n_{e}({\bf
        q})q^0\Big\{-{\rho}_{21}\Big[\frac{4id^{\nu\mu}}{q.k}{\epsilon^{\alpha\beta\rho}}_{\nu}
        q_{\mu}q_{\beta}({\epsilon}^1_{\alpha}{\epsilon}^1_{\rho}-
        {\epsilon}^2_{\alpha}{\epsilon}^2_{\rho})]+(\rho_{11}-\rho_{22})\no\\&& \!\!\!\!\!\!\!\!\!\!\!\!
        [\frac{d^{\nu\mu}}{(q.k)^2}({\epsilon}^1_{\beta}{\epsilon}^2
        _{\sigma}-{\epsilon}^1_{\sigma}{\epsilon}^2_{\beta})
        [(-2{q}_{\alpha}{k}_{\rho}({q}_{\mu}{q}_{\nu}
        +{k}_{\mu}{k}_{\nu})-2{q}_{\alpha}{q}_{\rho}({q}_{\mu}{k}_{\nu}+{k}_{\mu}{q}{\nu}))
        {\epsilon}^{\alpha\beta\rho\sigma}\no\\&& \!\!\!\!\!\!\!\!\!\!\!\!+(k^2{q}_{\alpha}{k}_{\mu}-2m^2{q}_
        {\mu}{k}_{\alpha}+2(q.k)q_{\alpha}q_{\mu}){{\epsilon^{\alpha\beta}}_{\nu}}^{\rho}]+
        \frac{2d^{\nu\mu}}{q.k}{\epsilon^{\alpha\beta\rho}}_{\nu}\no\\&& \!\!\!\!\!\!\!\!\!\!\!\!
        [q_{\alpha}{q}_{\rho}({\epsilon}^1_{\mu}{\epsilon}^2_{\beta}-{\epsilon}^2_{\mu}
        {\epsilon}^1_{\beta})+q_{\alpha}q_{\beta}({\epsilon}^2_{\mu}{\epsilon}^1_{\rho}-
        {\epsilon}^1_{\mu}{\epsilon}^2_{\rho})+2iq_{\mu}q_{\beta}{\epsilon}^1_{\alpha}
        {\epsilon}^2_{\rho}]\Big]\Big\}.
                \eea

\section{The calculation of density matrix elements for noncommutative theory}
The density matrix elements for the noncommutative theory are calculated in this appendix.
  In the Boltzmann equation (\ref{gb0}),
the commutator $[\mathcal{\hat H}_{int}(0),\mathcal{D}^{0}_{ij}]$ has the form
\bea
[\mathcal{\hat H}_{int}(0),\mathcal{D}_{ij}(\textbf{k})]\!\!\!\!\!\!\!\!&&=\int
d\textbf{q}d\textbf{q}'d\textbf{p}d\textbf{p}'(2\pi)^{3}\delta^{3}
(\textbf{q}'+\textbf{p}'-\textbf{q}-\textbf{p})(\mathcal{M}^{\theta}_1+\mathcal{M}^{\theta}_2+\mathcal{M}^{\theta}_3
+\mathcal{M}^{\theta}_4)\nonumber\\&&
\times[b^{\dag}_{r'}(q')b_{r}(q_1)a^{\dag}_{s'}(p')a_j(k)2p'^0(2\pi)^{3}\d_{is}\d^{3}(\textbf{p}-\textbf{k})
\nonumber\\&&\:\:\:\:\:\:\:\:\:\:\:\:\:\:\:\:\:\:\:\:\:\:
-b^{\dag}_{r'}(q')b_{r}(q)a^{\dag}_i(k)a_{s}(p)2p'^0(2\pi)^{3}\d_{js'}\d^{3}(\textbf{p}'-\textbf{k})],
\eea
where $\mathcal{M}^{\theta}_i$ are the noncommutative parameter
dependent parts of the amplitudes of the diagrams shown in Fig. 2. In
this figure the box shows vertex in the noncommutative space. These
amplitudes can be calculated straightforwardly as follows
\bea
\mathcal{M}^{\theta}\!\!\!\!\!\!\!\!&&=\mathcal{M}^{\theta}_1+\mathcal{M}^{\theta}_2+\mathcal{M}^{\theta}_3
+\mathcal{M}^{\theta}_4 \nonumber\\&&=\frac{-i\:e^2}{8p\cdot
q}\bar{u}_{r'}(q')\Big[\Big(\ep_{s'}\!\!\!\!\!\!/\:\:(p')q'\theta
(q+p)+q'\theta\ep_{s'}(p')p'\!\!\!\!/\:\Big)(q\!\!\!/+p\!\!\!/+m)\ep_{s}\!\!\!\!\!/\:\:(p)\no\\&&+
\ep_{s'}\!\!\!\!\!\!/\:\:(p')(q\!\!\!/+p\!\!\!/+m)\Big(\ep_{s}\!\!\!\!\!/\:\:(p)(q+p)\theta
q +q\theta\ep_{s}(p)p\!\!\!/\:\Big)\Big]u_{r}(q)\no\\&&
+\frac{i\:e^2}{8p'\cdot
q}\bar{u}_{r'}(q')\Big[\Big(\ep_{s}\!\!\!\!\!/\:\:(p)q'\theta
(q-p')-q'\theta\ep_{s}(p)p\!\!\!/\:\Big)(q\!\!\!/-p'\!\!\!\!/+m)\ep_{s'}\!\!\!\!\!\!/\:\:(p')\no\\&&+
\ep_{s}\!\!\!\!\!/\:\:(p)(q\!\!\!/-p'\!\!\!\!/+m)\Big(\ep_{s'}\!\!\!\!\!\!/\:\:(p')(q-p')\theta
q -q\theta\ep_{s'}(p')p\!\!\!/\:\Big)\Big]u_{r}(q), \eea
where $p\theta q\equiv p_{\mu}\theta^{\mu\nu}q_{\nu}$. Therefore one finds \bea
i\left<[\mathcal{\hat H}_{int}(0),\mathcal{D}_{ij}(\textbf{k})]\right>\!\!\!\!\!\!\!&&=\frac{e^2}{8m}\:\delta^{3}(0)\int
d\textbf{q}\:(2\pi)^3 \frac{n_e(\textbf{q})}{k\cdot
q}\:(\d_{is'}\rho_{s'j}(\textbf{k})-\d_{js'}\rho_{is}(\textbf{k}))\no\\&&
\!\!\!\!\!\!\!\!\!\!\!\!\!\!\!\!\!\!\!\!\!\!\!\!\!\!\!\!\!\!\!\!\!\!\!\!\!\!\!\!\!\!\!\!\!\!\!
 \times\bar{u}_{r'}(q)\Big[q\theta\ep_{s'}(k)\Big(k'\!\!\!\!/q\!\!\!/\ep_{s}\!\!\!\!\!/\:\:(k)+
 \ep_{s}\!\!\!\!\!/\:\:(k)q\!\!\!/k\!\!\!/\Big)+
q\theta\ep_{s}(k)\Big(\ep_{s'}\!\!\!\!\!\!/\:\:(k)q\!\!\!/k\!\!\!/+
k\!\!\!/q\!\!\!/\ep_{s'}\!\!\!\!\!\!/\:\:(k)\Big)
\Big]u_{r}(q),\eea

\begin{figure}\vspace{-1cm}
\centerline{\epsfysize=1.3in\epsfxsize=5.3in\epsffile{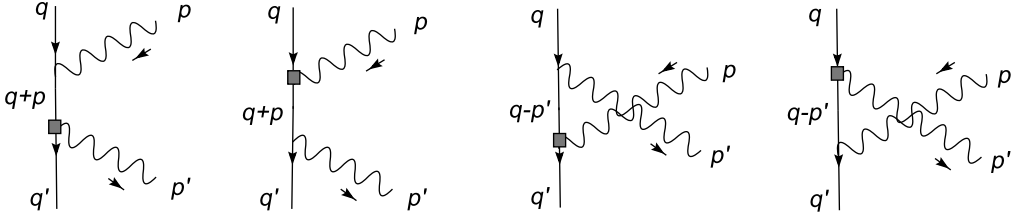}}\vspace{0.5cm}
\caption{Diagrams for noncommutative Compton scattering
 }\label{fig1}
\end{figure}

After some calculations the time evolution
of the density $\rho_{ij}$ generally is given by
\bea
2k^0\frac{d}{dt}\rho_{ij}(\textbf{k})\!\!\!\!\!\!&&=
\frac{e^2}{16m}\int d\textbf{q} \frac{n_e(\textbf{q})}{k\cdot
q}\:(\d_{is}\rho_{s'j}(\textbf{k})-\d_{js'}\rho_{is}(\textbf{k}))\no\\&&\!\!\!\!\!\!\!\!\!\!\!\!\!\!\!\!\!\!\!\!\!\!\!\!
\times\Big(q\theta\ep_{s'}(k)(16k\cdot
qq\cdot\ep_{s})+q\theta\ep_{s}(k)( 16k\cdot
qq\cdot\ep_{s'})\Big),\eea
where for the special cases $\dot{\rho}_{12}$ and
$\dot{\rho}_{21}$ one obtains
\bea \dot{\rho}_{12}(\textbf{k})\!\!\!\!\!\!&&=
\frac{e^2}{2mk^0}\int d\textbf{q}
\:n_e(\textbf{q})\Big(2(q\theta\ep_1q\cdot\ep_1+q\theta\ep_2q\cdot\ep_2)\rho_{12}-
(q\theta\ep_2q\cdot\ep_1+q\theta\ep_1q\cdot\ep_2)\rho_{11}\no\\&&
\:\:\:\:\:\:\:\:\:\:\:\:\:\:\:\:\:\:\:\:\:\:\:\:\:\:\:\:\:\:\:\:\:\:\:\:\:\:\:\:\:\:+(q\theta\ep_2q\cdot\ep_1
+q\theta\ep_1q\cdot\ep_2)\rho_{22}\Big)\label{ro1}, \eea
 and
  \bea
\dot{\rho}_{21}(\textbf{k})\!\!\!\!\!\!&&=-
\frac{e^2}{2mk^0}\int d\textbf{q}
\:n_e(\textbf{q})\Big(2(q\theta\ep_1q\cdot\ep_1+q\theta\ep_2q\cdot\ep_2)\rho_{21}-
(q\theta\ep_2q\cdot\ep_1+q\theta\ep_1q\cdot\ep_2)\rho_{11}\no\\&&
\:\:\:\:\:\:\:\:\:\:\:\:\:\:\:\:\:\:\:\:\:\:\:\:\:\:\:\:\:\:\:\:\:\:\:\:\:\:\:\:\:\:+(q\theta\ep_2q\cdot\ep_1
+q\theta\ep_1q\cdot\ep_2)\rho_{22}\Big)\label{ro2}, \eea
%%%%%%%%%%%%%%%%%%%%%%%%%%%%%%%%%%%%%%%%%%%%%%%%%%%%%%%%%%%%%%%%%%%%%%%%%%%%%%%%%

\section{The calculation of density matrix elements for magnetic field background}
In this appendix we calculate the density matrix elements in the presence of a background magnetic field.
Vanishing of the refractive term on the right hand side of (\ref{gb0}) forces us to go to the next order and
calculate the damping term of the Boltzmann equation (\ref{gb0}).
As it has been clarified in \cite{Kosowsky:1994cy}, using the
expectation values of creation and annihilation operators
calculated in this reference, the damping term can be written in
terms of amplitudes in the following form
\begin{eqnarray}\label{m15}
\frac{1}{2}\int_{-\infty}^{\infty}dt'&&\!\!\!\!\!\!\!\!\!\!
\langle[\hat{\mathcal{H}}_{int}(t'),[\hat{\mathcal{H}}_{int}(0),
\mathcal{D}_{ij}(\textbf{k})]]\rangle \nonumber \\&&
=\frac{1}{4}(2\pi)^{3}\delta^{3}(0)\int
d\textbf{q}d\textbf{p}d\textbf{q}'
\mathcal{M}(q'r',ps'_{1},qr,ks_{1})\mathcal{M}(qr,ks'_{2},q'r',ps_{2})\nonumber\\&&\:\:\:
\times(2\pi)^{4}\delta^{4}(q'+p-q-k)[n_{e}(\textbf{q})\delta_{s_{2}s'_{1}}
(\delta_{is_{1}}\rho_{s_{2}'j}(\textbf{k})+
\delta_{js_{2}'}\rho_{is_{1}}(\textbf{k}))\nonumber\\
&&\:\:\:\:\:\:\:\:\:\:\:-2n_{e}(\textbf{q}')\delta_{is_{1}}
\delta_{js_{2}'}\rho_{s_{1}'s_{2}}(\textbf{p})].
\end{eqnarray}
Up to the linear order of $B_{\alpha\beta}$, the evaluating of the
product of matrix elements yields
\begin{eqnarray}\label{m16}&&\!\!\!\!\!\!\!\!\!\!\!\!\!\!\!\!\!\!
\mathcal{M}(q'r',ps'_{1},qr,ks_{1})\times\mathcal{M}(qr,ks'_{2},q'r',ps_{2})
\mid_{\texttt{B-term}}\nonumber\\&&=-\frac{e^{4}}{4m^{2}}\Big(tr\{(q'\!\!\!\!/+m)
[\epsilon\!\!\!/_{s_{1}'}(p)S_{0}(q+k)
\epsilon\!\!\!/_{s_{1}}(k)+\epsilon\!\!\!/_{s_{1}}(k)S_{0}(q-p)\epsilon\!\!\!/_{s_{1}'}(p)]
(q\!\!\!/+m)\nonumber\\&&\:\:\:\:\:\:\:\:\:\:\times
[\epsilon\!\!\!/_{s'_{2}}(k)S_{B}(q'+p)
\epsilon\!\!\!/_{s_{2}}(p)+\epsilon\!\!\!/_{s_{2}}(p)S_{B}(q'-k)
\epsilon\!\!\!/_{s_{2}'}(k)]\}+tr\{(q'\!\!\!\!/+m)\nonumber\\
&&\:\:\:\:\:\:\:\:\:\:\times[\epsilon\!\!\!/_{s_{1}'}(p)S_{B}(q+k)
\epsilon\!\!\!/_{s_{1}}(k)+\epsilon\!\!\!/_{s_{1}}(k)S_{B}(q-p)\epsilon\!\!\!/_{s_{1}'}(p)]
(q\!\!\!/+m)[\epsilon\!\!\!/_{s'_{2}}(k)S_{0}(q'+p)
\epsilon\!\!\!/_{s_{2}}(p)\nonumber\\
&&\:\:\:\:\:\:\:\:\:+\:\epsilon\!\!\!/_{s_{2}}(p)S_{0}(q'-k)
\epsilon\!\!\!/_{s_{2}'}(k)]\}\Big),
\end{eqnarray}
where the \texttt{B-term} shows the terms that are linear in terms of the background magnetic field.
For relevant cosmological situations one can consider the cold plasma of electrons and photons that their kinetic energies are small compared to the electron mass i.e. $p, k$ and $q\ll m$.
As mentioned in \cite{Kosowsky:1994cy}, if the electron and photon temperatures are
comparable which means $p,k\ll q$, then various functions in (\ref{m15}) can be expanded in terms of $p/q$ and $q/m$.
Hence by computing the traces in (\ref{m16}), for the leading order,
the matrix element is simplified as
\begin{eqnarray}\label{m17}
\mathcal{M}(q'r',ps'_{1},qr,ks_{1})&&\!\!\!\!\!\!\!\!\!\times\mathcal{M}(qr,ks'_{2},q'r',ps_{2})
\mid_{\texttt{B-term}}\nonumber\\
&&\!\!\!\!\!\!\!\!=\frac{ie^{4}}{2m^{2}}\left(\frac{1}{q.k}-\frac{1}{q.p}\right)
\left(\frac{1}{(q.k)^{2}}-\frac{1}{(q.p)^{2}}\right)\nonumber\\
&&\!\!\!\!\!\!\!\!\times\Big[\tilde{q}.\epsilon_{s_{2}}(p)q.\epsilon_{s_{1}}(k)q.\epsilon_{s_{1}'}(p)q.\epsilon_{s_{2}'}(k)
-\tilde{q}.\epsilon_{s_{2}'}(k)q.\epsilon_{s_{2}}(p)q.\epsilon_{s_{1}}(k)q.\epsilon_{s_{1}'}(p)\nonumber\\
&&+\tilde{q}.\epsilon_{s_{1}}(k)q.\epsilon_{s_{1}'}(p)q.\epsilon_{s_{2}}(p)q.\epsilon_{s_{2}'}(k)
-\tilde{q}.\epsilon_{s_{1}'}(p)q.\epsilon_{s_{1}}(k)q.\epsilon_{s_{2}}(p)q.\epsilon_{s_{2}'}(k)
\nonumber\\&&\:\:\:\:\:\:\:\:\:+\mathcal{O}(k,p)\Big],
\end{eqnarray}
where $\tilde{q}_{\mu}=-eB_{\mu\nu}q^{\nu}$.
Now by substituting (\ref{m17}) in (\ref{m15}) and integrating over the $q'$, gives functions such as $n_{e}(\textbf{q}+(\textbf{k}-\textbf{p}))$ and $\delta(k-p+E(\textbf{q})-E(\textbf{q}+\textbf{k}-\textbf{p}))$ where can be expanded as follows
\begin{eqnarray}\label{m19}
n_{e}(\textbf{q}+(\textbf{k}-\textbf{p}))\sim n_{e}(\textbf{q})\left[1-\frac{(\textbf{k}-\textbf{p}).(\textbf{q}-m\textbf{v})}{mT_{e}}-\frac{(\textbf{k}-\textbf{p})^{2}}{2mT_{e}}
+\cdots\right],
\end{eqnarray}
and
\begin{eqnarray}\label{m20}
\delta(k-p+E(\textbf{q})-E(\textbf{q}+\textbf{k}-\textbf{p}))\sim
\delta(k-p)+\frac{(\textbf{k}-\textbf{p}).\textbf{q}}{m}
\frac{\partial\delta(k-p)}{\partial p}+\cdots.
\end{eqnarray}

Inserting the above equations  into the (\ref{m15}) leads to the density matrix $\rho_{12}$ and $\rho_{21}$.

%%%%%%%%%%%%%%%%%%%%%%%%%%%%%%%%%%%%%%%%%%%%%%%%%%%%%%%%%%%%%%%%%%%%%%%%%%%%%%%%%%

\end{document}